\documentclass[12pt]{article}
\pdfoutput=1

\usepackage{amsmath}
\usepackage{amsthm}
\usepackage{amscd}
\usepackage{amsfonts}
\usepackage[psamsfonts]{amssymb}
\usepackage{graphicx}
\usepackage{epsfig}

\newcommand\beq{\begin{eqnarray}}
\newcommand\eeq{\end{eqnarray}}

\begin{document}
\begin{titlepage}

\bigskip\bigskip

\begin{center}
{\Large{\bf Finite temperature Aging Holography}}

\bigskip\bigskip
Seungjoon Hyun$^{1}$, Jaehoon Jeong$^1$ and 
Bom Soo Kim$^{2,3}$
\bigskip

${}^1${\it {\small Department of Physics, College of Science, Yonsei University, Seoul 120-749, Korea}} \\
${}^2${\it {\small Crete Center for Theoretical Physics, University of Crete, Crete, Greece}} \\
${}^3${\it {\small IESL - FORTH, P.O.Box 1527, 71110 Heraklion, Crete, Greece}} \\

\end{center}

\bigskip

{\small
\centerline{sjhyun@yonsei.ac.kr, ~~j.jeong@yonsei.ac.kr, ~~bskim@physics.uoc.gr}
}

\bigskip\bigskip
\begin{abstract}
We construct the gravity background which describes the dual field theory with aging invariance. We choose the decay modes of the bulk scalar field in the internal spectator direction  to obtain the dissipative behavior of the boundary correlation functions of the dual scalar fields. In particular, the two-time correlation function at zero temperature has  the characteristic features of the aging system: power law decay, broken time translation and dynamical scaling. We also construct the black hole backgrounds with  asymptotic aging invariance. We extensively study characteristic behavior of the finite temperature  two-point correlation function via analytic and numerical methods. 
 
\end{abstract}

\vspace{0.5in}
\end{titlepage}

\tableofcontents

\section{Introduction}

AdS/CFT correspondence has been successful in describing strongly-coupled field theories 
using weakly-coupled classical gravity backgrounds, providing new paradigm of analytical methods 
for theories with strong coupling \cite{Maldacena:1997re}\cite{Aharony:1999ti}. 
Recently this correspondence extended its application to non-relativistic setup with 
general dynamical exponent z $\neq$ 1. Those non-relativistic holographic theories include 
Schr\"odinger \cite{Son:2008ye, Balasubramanian:2008dm, Goldberger:2008vg, Barbon:2008bg} 
and Lifshitz holography \cite{Kachru:2008yh}.%
	\footnote{See {\it e.g.} \cite{Horava:2009uw} for a very different usage of the Lifshitz symmetry 
	with z=3 in the context of quantum field theory of gravity.
	}
Since then, they have been enjoying continuous attractions. 

An acquaintance of non-relativistic holographic theories with z $\neq$ 1 tells that 
we can construct a plethora of explicit time dependent backgrounds by utilizing the distinguished 
role of time. This is different from the relativistic theories. Thus we might hope to  
shed lights on time dependent backgrounds in string theory and in AdS/CFT correspondence  
even though they are in general known to be challenging.  
We would like to add some contributions to this direction in the context of the 
non-relativistic Schr\"odinger holography with z=2.  

One of the simplest applications of the non-equilibrium physics is known as {\it aging}. 
Aging phenomena can be characterized by two-time correlation functions, 
which typically show that older systems relax in a slower manner than younger systems 
after entering a given physical phase under study, see {\it e.g.} \cite{Cugliandolo:2002dy}\cite{Helkel:2007nept} 
for reviews. Two pioneering works of aging in holographic setup have been put forward in  
\cite{Minic:08073665}\cite{Jottar:2010vp} by generalizing Schr\"odinger background 
\cite{Son:2008ye}\cite{Balasubramanian:2008dm} with explicit time dependent terms. 
The authors of \cite{Jottar:2010vp} observe that aging algebra can be realized 
as a subset of Schr\"odinger algebra using a singular time dependent coordinate transformation. 
See also \cite{Nakayama:2010xq} for similar observations of general time dependent deformations 
of Schr\"odinger backgrounds. 

Along the line of Schr\"odinger backgrounds \cite{Son:2008ye}\cite{Balasubramanian:2008dm}, 
there have been progresses with much interest. 
Their finite temperature generalizations with the thermodynamic analysis can be found in 
\cite{Herzog:2008wg, Maldacena:2008wh, Adams:2008wt, Yamada:2008if} and 
transport properties are analyzed in \cite{Ammon:2010eq}, recently. 
These solutions are constructed nicely using the null Melvin twist 
\cite{Gimon:2003xk}\cite{Alishahiha:2003ru}. There also have been efforts constructing 
Schr\"odinger solutions from string or M theory, sometimes with supersymmetry. See   
\cite{Hartnoll:2008rs,Duval:2008jg,Mazzucato:2008tr,Jeong:2009aa,Kim:2011fb}, for example.
While these Schr\"odinger backgrounds have been drawn much interest, 
the progresses directly along this line have been hindered due to the conceptual difficulties 
related to their non-trivial boundary structures. Furthermore 
one quickly finds that finite temperature generalizations of the Schr\"odinger backgrounds 
are complicated, and thus practical calculations are difficult. 

An alternative candidate for the Schr\"odinger holography, AdS in light-cone, has been constructed in 
\cite{Goldberger:2008vg}\cite{Barbon:2008bg} even before the finite temperature generalizations of the 
Schr\"odinger backgrounds. Its finite temperature generalizations were done 
in \cite{Maldacena:2008wh}\cite{Kim:2010tf} with thermodynamic analysis, and 
transport properties are also analyzed in \cite{Kim:2010tf}\cite{Kim:2010}.
It turns out that these two different holographic theories, Schr\"odinger backgrounds 
and AdS in light-cone, have the same thermodynamic properties \cite{Maldacena:2008wh}\cite{Kim:2010tf} 
and also the same transport properties when the comparisons are reliable \cite{Kim:2010tf}.   
Thus AdS in light-cone is a viable candidate for Schr\"odinger holography. 
While it remains as a question whether these two theories describe the same physical properties or not, 
AdS in light-cone has advantages over the Schr\"odinger backgrounds: 
it has well-defined holographic renormalization and is as simple as original AdS in its computations.    

Motivated by these observations, we would like to construct a holographic model of aging 
in the context of \cite{Goldberger:2008vg}\cite{Barbon:2008bg}, Aging in light-cone, 
by realizing the aging 
symmetry with the singular time dependent coordinate transformation. This is done in section 
\ref{sec:ZeroTemp}. From the bulk gravity side, we obtain the two-point correlation function of the dual field theory.
In particular, we consider the, so-called, spectator coordinate $x^-$ as an {\it internal} direction.%
	\footnote{
	Note that the geometric realizations of Schr\"odinger symmetry separates itself from the other holographic examples, 
	providing a correspondence between $(d,1)$-dimensional field theory and $(d+3)$-dimensional gravity theory. 
	In addition to the usual radial direction, it has another coordinate $x^-$, which does not contribute to the coordinates 
	in field theory. The $x^-$ coordinate is known to provide a non-relativistic particle number or mass, referred here by $M$. 
	We call this spectator direction, $x^-$, as {\it internal} because it does not show up explicitly in the dual field theory. 
		
	While the full potential associated with the $x^-$ direction remains to be seen, 
	we believe that the treatment adapted in this paper is one of the first non-trivial examples of utilizing this spectator direction. 
	See {\it e.g.} \cite{Guica:2010sw}\cite{Balasubramanian:2010uw} for different attempts on this issue.
	}
Furthermore, since we are dealing with time dependent and dissipative system, we choose the decay modes of the bulk scalar field in the internal spectator direction. This is in contrast with  the complexification of spacetime done in \cite{Jottar:2010vp}. As a result, we obtain generalized two-time correlation function of the dual field theory, which exhibit the characteristic features of the aging system. 

In section 3 we consider finite temperature generalizations of the aging holography on  Aging in light-cone  
as well as Aging in Schr\"odinger backgrounds. We focus on the former case, 
which has simpler form, and then give some comments on the latter. We study the finite temperature two-point correlation function 
in Aging in light-cone via AdS/CFT correspondence. We provide the novel time dependent structure of aging two-point 
correlation function in coordinate space, compared to that of Schr\"odinger case and to that of momentum space.     
As the latter is not exactly solvable, we use numerical as well as analytic methods to see its behavior. 
Finally, we give some comments on the aging phenomena from the Schr\"odinger background stressing the difference 
compared to those of the Aging in light-cone. It turns out that the internal isometry direction 
plays a crucial role throughout in this paper. In section 4, we give conclusion.

\section{Zero Temperature Aging Holography} \label{sec:ZeroTemp}

In this section, we start with a brief review of aging holography based on 
recent papers \cite{Minic:08073665}\cite{Jottar:2010vp} 
and also fix some notations. The relevant geometry is simple and possesses 
lots of isometries, yet there are many difficulties in its fundamental level. 
We also notice \cite{Nakayama:2010xq}, 
which considered some universal time dependent deformations in the context of Schr\"odinger 
geometry. One of the deformations is essentially the same as that of \cite{Jottar:2010vp}, yet 
the discussion was brief and not developed to the degree done in \cite{Jottar:2010vp}.

Then we develop the basic two-point correlation functions from a background called 
AdS in light-cone \cite{Goldberger:2008vg,Barbon:2008bg,Maldacena:2008wh,Kim:2010tf} 
following \cite{Minic:08073665}\cite{Jottar:2010vp}. On the way, we develop several conceptually important 
points of the applications of AdS/CFT on time dependent backgrounds. They include \\
i) Establishing the holographic aging equation which describes non-equilibrium critical phenomena,  \\
ii) Non-trivial radial fall offs of wave function, which are different from the conformal dimensions 
of the corresponding field, and  \\
iii) Utilizing the $x^-$ coordinate as an internal direction from the point of view of holography, 
which gives an alternative way to get real two-time correlators instead of complexifying the geometry 
done in \cite{Jottar:2010vp}. 

To make a clear distinction between these two approaches, 
we refer Aging background and Schr\"odinger background for the geometric realization with the 
additional terms in the background \cite{Son:2008ye}\cite{Balasubramanian:2008dm}\cite{Jottar:2010vp}, 
while we use the term Aging in light-cone and AdS in light-cone for  \cite{Goldberger:2008vg}\cite{Barbon:2008bg}\cite{Kim:2010tf}.

\subsection{Zero Temperature Aging Background : A Review}\label{sec:AgingBac} 

Starting with the Schr\"odinger background \cite{Son:2008ye}\cite{Balasubramanian:2008dm}
\begin{align}
		ds^2 &= r^2 \left(  d\vec{y}^2 - 2 dx^{+} dx^{-} -\tilde \beta^2 r^2 d x^{+2} \right) + \frac{dr^2}{r^2} \;, 
		\label{SchBackrMet}
\end{align}
the following Aging background is constructed in \cite{Jottar:2010vp} 
\begin{align}
		ds^2 &= r^2 \left[  d\vec{y}^2 - 2 dx^{+} dx^{-} - \left( \tilde \beta^2 r^2 +  \frac{\tilde \alpha \tilde \beta}{x^+} \right) d x^{+2} \right] -2 \tilde \alpha \tilde \beta  r dr dx^++ \frac{dr^2}{r^2} \;. 
		\label{AgingBacMet}
\end{align}
This turns out to be one of the simplest time dependent background which possesses a large number of isometries called 
aging symmetry with dynamical exponent z=2. The Dilatation scaling symmetry $D$ is associated with 
the transformations, $\vec y \rightarrow \lambda \vec y$, 
$x^+ \rightarrow \lambda^2 x^+$ and $x^- \rightarrow x^-$. Other relevant symmetries are spatial translations 
$P_i$, Galilean boosts $K_i$, rotations $R_{ij}$, central element $M$ interpreted as a particle number and 
the special conformal transformation $C$, while the time 
translational symmetry is broken by the explicit time dependence of the metric {\it globally}. We come back to this below.

There exist several valuable observations in \cite{Jottar:2010vp}, which are related to the geometry in
equation (\ref{AgingBacMet}) and worthwhile to be mentioned: 
\begin{itemize}
\item The metric (\ref{AgingBacMet}) is connected to the Schr\"odinger background
	 (\ref{SchBackrMet}) by a local, but {\it singular}, coordinate transformation 
		\begin{equation}
		x^- \quad \longrightarrow \quad  x^- - \frac{\tilde \alpha \tilde \beta}{2}  \ln \left( r^2 x^+ \right) \;.
		\label{CoordinateChange}
		\end{equation}
	This singular time dependent coordinate change is a key to realize aging symmetry.
	It does not generate curvature singularities at any point in spacetime.
	It is argued in \cite{Jottar:2010vp} that this coordinate singularity in this context 
	is similar to a black hole horizon, where physical boundary conditions are imposed, and thus has profound effects,   
	from the gauge/gravity point of view. This geometry is called {\it locally Schr\"odinger} in \cite{Jottar:2010vp}. 
	Thus it is important to impose physical boundary conditions on the time boundaries in addition to the 
	spatial boundaries.
	
\item Thus, locally, the Aging background has the full Schr\"odinger symmetry with modified set of generators, especially 
	the time translation ($H_A$) and special conformal transformation ($C_A$) generators as 
		\begin{align}
			H_A &= \partial_{x^+} - \frac{\tilde \alpha \tilde \beta}{2 x^+ } \partial_{x^-} \;, \nonumber \\
			C_A &= - x^+ r \partial_r + x^+ \vec y \cdot \vec \partial + x^{+2} \partial_{x^+}
			+ \frac{1}{2} \left( \vec y^2 + \frac{1}{r^2} - \tilde \alpha \tilde \beta x^+ \right) \partial_{x^-} \;.
			\label{ModifiedIsom}
		\end{align}
	The other relevant symmetry generators are not changed by the coordinate transformation (\ref{CoordinateChange})
	\cite{Jottar:2010vp}.	Globally the time translation symmetry is broken and 
	the aging symmetry is realized as conformal Schr\"odinger symmetry modulo time translation, and described by the generators 
	$\{R_{ij}, P_i, K_i, D, C, M \}$. 

\item The analysis is done with the complex $x^+$, especially at $x^+ = 0$, and also the geometry is extended to be 
	complex with complexified parameter $\tilde \alpha$ in \cite{Jottar:2010vp}. 
	With real $\tilde \alpha$, the resulting correlator 
	has the time dependence only in its phase. To achieve a relaxation process, $\tilde \alpha$ should be complex.   

	We show that there is a way out from complexifying the geometry by utilizing the fact that 
	$x^-$ can be considered as an {\it internal} coordinate from the holographic point of view.
\end{itemize}

\subsection{Aging in Light-Cone}\label{sec:AgingLC}

In a similar manner, we start with the AdS in light-cone \cite{Goldberger:2008vg,Barbon:2008bg}
\begin{align}
		ds^2 &= r^2 \left(  d\vec{y}^2 - 2 dx^{+} dx^{-} \right) + \frac{dr^2}{r^2} \;, 
		\label{AdSLCMet}
\end{align}
which is shown to be another viable candidate for the Schr\"odinger holography. 
This metric can be obtained from the AdS metric with the following coordinate change 
\begin{align} 
	x^+ = b(t+x)   \;,\qquad    x^- = \frac{1}{2b}(t-x)   \;.
	\label{LightConeCoordinate}
\end{align}	
This coordinate transformation was introduced in \cite{Maldacena:2008wh}\cite{Kim:2010tf}\cite{Kim:2010} 
and should be viewed as a two-step procedure:
a boost in the $x$-direction with rapidity $\log b$,
followed by transforming to light-cone coordinates.
To ensure z=2, we assign $[b]$ (the scaling dimension of $b$ in the unit of mass) as $-1$, and thus
$[y_i]=-1$, $[x^+] = -2$ and $[x^-]= 0$. To achieve the desired Schr\"odinger 
isometry, it is further required to identify the coordinate $x^+$ as time and also to have a momentum 
projection along the other coordinate $x^-$. $x^-$ direction provides an isometry to the background and 
thus the corresponding momentum gives the central element $M$ which  serves as the
total particle number or the mass of the resultant non-relativistic theory.   
Furthermore this AdS in light-cone is known to have a well defined holographic renormalization 
and as simple as the case of AdS black holes \cite{Kim:2010tf}.%
	\footnote{See {\it e.g.} \cite{Balasubramanian:1999re}\cite{de Haro:2000xn} for holographic 
	renormalization in general, \cite{Ross:2009ar} for Schr\"odinger background using 
	modified definition of the stress tensor, and \cite{Horava:2009vy} for general anisotropic 
	backgrounds using anisotropic scaling.
	} 

It has been known that the Schr\"odinger background and the AdS in light-cone share the same 
physical properties such as thermodynamic \cite{Maldacena:2008wh,Kim:2010tf} 
and transport properties \cite{Kim:2010tf}. 
We would like to investigate whether this similarity extends to this time dependent settings or not.   

With the {\it singular} coordinate transformation (\ref{CoordinateChange}) and a notation change 
$\tilde \alpha \tilde \beta ~\rightarrow ~ \alpha$, we obtain the following metric
\begin{align}
		ds^2 &= r^2 \left(  d\vec{y}^2 - 2 dx^{+} dx^{-}  -  \frac{\alpha}{x^+} ~dx^{+ 2} \right)
		-2 \alpha ~  r dr dx^+ + \frac{dr^2}{r^2} \;, 
		\label{AgingLCMetric}
\end{align}
which is simpler than that of the Aging background (\ref{AgingBacMet}). One may note that  $\alpha$ is a dimensionless parameter.
It can be shown that this metric shares the similar properties as the geometry  (\ref{AgingBacMet}), mentioned in the previous section 
\ref{sec:AgingBac}. First, the metric (\ref{AgingLCMetric}) is locally AdS in light-cone, while time translation symmetry is broken 
globally by the {\it singular} 
coordinate transformation which connects these two metrics (\ref{AgingLCMetric}) and (\ref{AdSLCMet}). 
Second, the isometry of the metric (\ref{AgingLCMetric}) is aging symmetry with the same set of the 
generators given in  (\ref{ModifiedIsom}).  
 
With these properties in mind, let us investigate the background (\ref{AgingLCMetric}). For this purpose, we change the coordinate as $u=\frac{L^2}{r}$  
and concentrate on the five dimensional background with $d \vec y^2 = dy_1^2 + dy_2^2$ and
the metric becomes
\begin{eqnarray}
	ds_u^2 =   \frac{L^2}{u^2} \left(  dy_1^2 + dy_2^2  -2 dx^{+}  dx^{-} - \frac{\alpha}{x^+} ~dx^{+ 2}  
		+ \frac{2 \alpha  }{u} du dx^+  +  du^2  \right) \ ,
\end{eqnarray}
where the boundary sits at $u=0$.
Let us couple the background with a probe scalar $\phi$, whose action has the following form 
\begin{align}
	S = K \int d^4 x \int_{u_B}^{\infty} du \sqrt{-g}
	& \left(  g^{uu} ~\partial_u \bar \phi ~\partial_u \phi  +  g^{u x^-} ~\partial_u \bar \phi ~\partial_{x^-} \phi \right. \nonumber \\
 	& \left.
	+ g^{u x^-} ~\partial_{x^-} \bar \phi ~\partial_u \phi
	+ g^{\mu\nu} ~\partial_{\mu} \bar \phi ~\partial_{\nu} \phi + m^2 ~\bar \phi ~\phi \right)  \ , 
	\label{ScalarAction}
\end{align}
where $K = -\pi^3 L^5 /4 \kappa_{10}^2$ and $\mu, \nu = +, -, y, z$.
We used $u_B$ for the boundary cutoff, which is small and will be sent to zero eventually. 
The linearized field equation for $\phi$ becomes
\begin{align}
2M \left(i  \frac{\partial}{\partial x^+}  +  \frac{\alpha  M}{2 x^+}   \right) \phi &  \nonumber \\
=  \frac{\partial^2 \phi}{\partial u^2} &+ (2 i M\alpha -3) \frac{1}{u} \frac{\partial \phi}{\partial u} 
-\left ( \frac{4i M\alpha + \alpha^2 M^2 + m^2 L^2 }{u^2} + \vec \nabla ^2 \right) \phi
  \;.
\label{BlukScalarEq}
\end{align}
Note that here we treat $x^-$ coordinate special and replace all the $\partial_{x^-}$ as $iM$, 
because this coordinate plays a distinguished role in the Sch\"ordinger holography. 
Later on we consider the parameter $M$ to be either {\it real} or {\it imaginary}, from the observation that 
the coordinate $x^-$ can be treated as an internal coordinate from the field theory point 
of view.

\subsection{Two-point Correlation Function}\label{ZeroTCorrelationFunction}

To find the solution of the equation (\ref{BlukScalarEq}), we use the Fourier decomposition as
\begin{eqnarray}
\phi(u,x^+, \vec y) = \int \frac{d \omega}{2 \pi} \frac{d^2 k}{(2\pi )^2} ~e^{i \vec{k} \cdot \vec{y}} 
~T_{\omega}(x^{+})~ f_{\omega,\vec{k}}(u) ~\phi_0 (\omega,\vec{k}) \ ,
\label{IntTran}
\end{eqnarray}
where $\vec{k} $ is the momentum vector for the corresponding coordinates $\vec y $.
$\phi_0(\omega,\vec{k}) $ is introduced for the calculation of the correlation functions and 
is determined by the boundary condition with the normalization $f_{\omega,\vec{k}}(u_B) =1$. 
And $T_{\omega}(x^{+})$ is the kernel of integral transformation that convert $\omega$ to $x^{+}$, 
which is necessary for our time dependent setup. 

With this Fourier mode, the differential equation (\ref{BlukScalarEq}) decomposes into 
time dependent part and radial coordinate dependent part. The time dependent equation and its solution read    
\begin{eqnarray}
 \left( \frac{\alpha  M}{2 x^+} +  i  \partial_{+} \right) T_{\omega}   = \omega T_{\omega} \qquad \longrightarrow \qquad
T_{\omega}(x^+) 
=  c_1  \exp^{-i \omega x^+} (x^+)^{\frac{i\alpha M}{2}} \;.
\end{eqnarray}
The radial dependent equation is given by
\begin{align}
u^2 f_{\omega,\vec{k}} '' +(2 i M\alpha -3) u f_{\omega,\vec{k}}' 
-\left ( 4i M\alpha + \alpha^2 M^2 + m^2 L^2 + u^2 \vec k^2 \right) f_{\omega,\vec{k}}
= 2 M\omega ~u^2 f_{\omega,\vec{k}} \;, 
\label{radial equation}
\end{align}
where $f^\prime=\partial_u f$. An analytic solution is available in terms of Bessel functions as 
\begin{align}
f_{\omega,\vec{k}} =u^{2-i \alpha M} \left(  c_2 I_\nu (q u) + c_3  K_\nu (q u) \right) \;,
\end{align}
where 
$I_\nu$ and $K_\nu$ are Bessel functions with $\nu= \sqrt{4+L^2 m^2 } $ and $q =   \sqrt{\vec k^2 +2 M \omega}$.

For this solution to be well defined, we need to impose two different physical boundary conditions, 
one in deep inside the bulk and another at $x^+ =0$, because we consider only $x^+ \geq 0$. 
To have a well defined field deep in the bulk, we choose $K$ over $I$,
\begin{eqnarray}
f_{\omega,\vec{k}} = c ~ u^{2-i\alpha M}  K_\nu (q u)\;,
\end{eqnarray}
which is exponentially converging for large $u$, deep in the bulk. 
Thus the full solution is
\begin{align}
\phi(u,x^+, \vec y) = \int \frac{d^2 k}{(2\pi )^2}\frac{d \omega}{2 \pi} e^{i \vec{k} \cdot \vec{y} -i \omega x^+} u^2  \left( \frac{\alpha x^+}{u^2} \right)^{\frac{i\alpha M}{2}} ~c~ K_\nu (q u) ~ \phi_0 (\omega,\vec{k}) \;. 
\label{AgingWaveFunction}
\end{align}
Thus we confirm that the aging scalar function is scale-invariant prefactor 
$\left( \frac{\alpha x^+}{u^2} \right)^{\frac{i\alpha M}{2}} $ times that of the Schr\"odinger background, advertised 
in \cite{Jottar:2010vp} 
\begin{align}
\phi_{\text{Aging}}(u,x^+, \vec y) =  \left( \frac{\alpha x^+}{u^2} \right)^{\frac{i\alpha M}{2}} \phi_{{ \text{Schr\"odinger}}}(u,x^+, \vec y) \;. 
\label{FinalWaveFunction}
\end{align}

We consider the $x^-$ coordinate as the one for the internal manifold and restrict the system to be  a sector with an eigenvalue $M$. The parameter $M$ would be real if the $x^-$ coordinate  is periodic as in discrete light cone quantization. We assume that  the $x^-$ has boundaries, for example, at $x^-=0$ and take $M$ to be a general complex number. This seems to accord with the the boundary at $x^+=0$. It would be natural to expect that the imaginary $M$ would represent the dissipative behavior of the system.
Let us consider the radial fall offs of the wave solution of the scalar field with the conformal dimension $2-\nu$.
At the boundary $u \rightarrow 0$, the solution behaves as
\begin{align}
f_{\omega,\vec{k}} 
\sim u^{2-\nu}  u^{-i\alpha M} \sim  u^{2-\nu + \alpha M_I}  u^{-i\alpha M_R} \;, \quad \text{with} \quad 
M = M_R + i M_I \;.
\end{align} 
Note that we consider complex $M = M_R + i M_I$ for notational simplicity. 
Below, we will consider real $M=M_R$ and imaginary $M=i M_I$ cases separately to make the distinction clear.

It is worthwhile to pause and have some comments. First, the conformal dimension of the scalar field 
$\phi$ is the same as in AdS in light-cone because the factor $\left( \frac{\alpha x^+}{u^2} \right)^{\frac{i\alpha M}{2}} $ 
is scale invariant and thus does not contribute to the scaling dimension, which is observed in \cite{Jottar:2010vp}. 
Second, as a result, the fall-off behavior of the scalar wave solution at the boundary changes for the case $M=i M_I$.
This explicitly demonstrates that the correct scaling dimension of some 
operators can not be read off just from the behavior of the radial wave solution in the time dependent holography. 
Third, this change of the radial wave solution can be viewed as "wave function renormalization" 
and should not be canceled by additional counter terms, as observed in \cite{Jottar:2010vp}. 
Physical quantities would have appropriate time dependence to compensate this extra effect.  
This will be explicitly demonstrated by evaluating the correlation functions  below. 

We further require the scalar wave solution to behave well as it approaches to $x^+ \rightarrow 0$, 
because $x^+ =0$ also serves as a boundary. 
This is similar to the boundary condition imposed deep inside the bulk along the $u$-direction, which 
leads physically reasonable results below. To check the behaviors, we calculate the wave solution, 
equation (\ref{AgingWaveFunction}), near the boundary in coordinate space for the massless case $m=0$ with 
$M=i M_I$, $\nu=2$ and $\phi_0 (\omega,\vec{k})=1$ as
\begin{align}
\theta(x^+) (x^+)^{-\frac{\alpha M_I}{2}-1} \exp \left\{-\frac{ M_I \vec y^2}{2 x^+}\right\} \;.
\end{align}
This shows the typical behaviors of the wave solution in the coordinate space. One sees 
that the wave solution converges at $x^+ =0$ for $M_I >0$ due to the exponential factor and also 
converges at $x^+ \rightarrow \infty$ for $\frac{\alpha M_I}{2}+ 1>0 $ due to the polynomial factor. 
This demonstrates the convergence of the wave solution at the time boundaries for the imaginary $M=i M_I$. 
From the fact that the asymptotic forms of the wave solution and two point correlation function are restricted by aging symmetry, 
one expects to get similar exponential and polynomial factors for more general cases.

We follow \cite{Son:2002sd}\cite{Son:2007vk}\cite{Skenderis:2008dh}\cite{Skenderis:2008dg} 
to compute the correlation functions by introducing a cutoff $u_B$ near the boundary and 
normalizing $f_{\omega,\vec{k}}(u_B)  = 1$, which fixes $ c =  u_B^{-2+ i \alpha M} K_\nu^{-1} (q u_B)$. 
Let us consider the on-shell action, which has two different contributions, the terms proportional to $g^{uu}$ and 
to $g^{u-}$, as 
\begin{align}
	S[\phi_0] 
	&= \int d^3 x  \frac{L^5}{u^5}  ~\phi^* (u,x^+, \vec y) ~\left(\frac{u^2}{L^2}\partial_u + i M \frac{\alpha u}{L^2}  \right) \phi (u,x^+, \vec y) \big |_{u_B} \;. 
\end{align}
This can be recast using the equation (\ref{AgingWaveFunction}) as 
\begin{align}	
	&\int d x^+  ~\theta (x^+) ~\frac{d \omega'}{2 \pi} \frac{d \omega}{2 \pi} e^{-i (\omega' -\omega) x^+} 
	\left( \alpha x^+ \right)^{-\frac{i\alpha (M^*-M)}{2}} 
	\nonumber \\
	&\qquad \times 
	\int d^2 y  \int \frac{d^2 k'}{(2\pi)^2} \int \frac{d^2 k}{(2\pi)^2} 
	e^{i(\vec k' - \vec k) \cdot \vec y} ~ \phi_0^* (\omega',\vec{k'}) {\cal F}(u, \omega', \omega, \vec{k'}, \vec{k}) \phi_0
	(\omega,\vec{k}) \big |_{u_B} \;, 
	\label{wholeEQ}
\end{align}
where $\theta (x^+)$ represents the existence of the physical boundary along the time direction as $0 \leq x^+ < \infty$, 
and ${\cal F}$ is given by 
\begin{align}
	{\cal F} (u, \omega', \omega, \vec{k'}, \vec{k})
	&= \frac{L^5}{u^5} f_{\omega',\vec{k'}}^* (\omega',\vec{k'},u) \left(\frac{u^2}{L^2}\partial_u
	+ i M \frac{\alpha u}{L^2}  \right) f_{\omega,\vec{k}} (\omega,\vec{k},u) .
\end{align}
Note that the spatial integration along $\vec y$ can be done to give delta function $\delta^2 (\vec k' -\vec k)$. 
One can bring the $u^{\pm i \alpha M}$ factors in $f$ and $f^*$ together to cancel each other, which
removes the second part in $ {\cal F}$. Then using the relation 
\begin{align}
\frac{\partial}{\partial u}  u^2   K_\nu (q u) = u \{ (2-\nu)  K_\nu (q u) - q u K_{\nu-1}  (q u) \} \;,
\end{align}
one can evaluate the $u$-dependent part at $u=u_B$ at the boundary by 
expanding in terms of small $u_B$ to obtain the non-trivial contribution in ${\cal F}$ as
\begin{align}
	{\cal F}(u_{B},\omega,\vec{k})
	&=  - \frac{2 \Gamma (1-\nu)}{ \Gamma (\nu)}
	\left( \frac{L^3}{u_B^4} \right)    \left( \frac{q u_B}{2}\right)^{2\nu} + \cdots   \;.
	\label{FFfunction}
\end{align}
Note that the function ${\cal F}$ is only function of $\omega$ and $\vec k$ when it is evaluated at the boundary. 
Thus the time independent part is the same as previously considered cases in \cite{Balasubramanian:2008dm}
\cite{Goldberger:2008vg}\cite{Jottar:2010vp}. But this is not the end of the story. 
To obtain the final form of the momentum correlation function, we need to evaluate the time dependent part. 
Here we consider two cases, $M=M_R$ and $M=i M_I$ with real parameters $M_R$ and $M_I$, separately.

 \subsubsection{$M=M_R$}

One can perform the $x^+$ integration in equation (\ref{wholeEQ}) to get the following results,  
because of the step function and vanishing exponent, $M^* - M =0$, 
\begin{align}
{\cal G}  (\omega' - \omega) = \frac{\delta (\omega' - \omega)}{2}  -\frac{i}{2 \pi  (\omega' - \omega)} \;.
\label{GMR}
\end{align}
Thus the momentum correlation function has two parts  
\begin{align}
	&\langle {\cal O}^* (\omega',\vec{k}') {\cal O}(\omega, \vec{k}) \rangle
	= -2 (2\pi)^{-3}  \delta (\vec{k}' - \vec{k} ) ~ {\cal F}(u_{B},\omega,\vec{k})~ {\cal G}  (\omega' - \omega) 
	\;,
\end{align}
where ${\cal F}$ is given in equation (\ref{FFfunction}). 
This is the final form of the momentum space correlation functions. Note the extra term present due to our 
boundary condition with step function in time domain. 

Let us evaluate the coordinate correlation function as%
\footnote{Note the time dependent factor $\left( x^+\right)^{-i \frac{\alpha M_R}{2}}$ 
in front of the boundary scalar operator is opposite to that of the corresponding scalar field in (\ref{AgingWaveFunction}). 
The scaling dimension of the aging scalar operator is effectively different by $i \alpha M$ compared to that of 
the Schr\"odinger case 
\begin{align}
[\phi_{\rm aging}] = [\phi_{\text{Schr\"odinger}}] +  i \alpha M \;,
\end{align}
as read off from the radial fall-offs in equation (\ref{FinalWaveFunction}). 
This should be compensated by an appropriate time dependent factor.  
Another way to see this is the the relation 
\begin{align}
\langle {\cal O} (1) \cdots \rangle = \frac{\delta}{\delta \phi_0 (1)} \cdots  e^{\int_{\partial} {\cal O} \phi_0 } \;, \nonumber 
\end{align} 
where the time dependent factor would result in opposite way. This is the same for the case with 
$M=i M_I$ discussed below.
}  
\begin{align} 
&\langle {\mathcal O}^* (x_{2}^{+},\vec{y}_{2})  {\mathcal O}(x_{1}^{+},\vec{y}_{1}) \rangle \nonumber \\
&\quad = \int \frac{d \omega'}{2 \pi} \frac{d^2 k'}{(2\pi )^2} \frac{d \omega}{2 \pi} \frac{d^2 k}{(2\pi )^2} 
e^{i \vec{k'} \cdot \vec{y}_{2} -i \vec k \cdot \vec{y}_{1} } e^{ -i \omega' \cdot x^{+}_{2} + i \omega \cdot x_1^+} 
\left(\frac{ x_2^+}{ x_1^+}\right)^{i \frac{\alpha M_R}{2}}
\langle {\cal O}^* (\omega',\vec{k}') {\cal O}(\omega, \vec{k}) \rangle \;. 
\end{align} 
For further calculations, we use the following integral
\begin{align}
\int \frac{d \omega'}{2 \pi} e^{- i (\omega' -\omega) x_2^+ } {\cal G}  (\omega' - \omega)  = \frac{1+ \text{sign} (x_2^+) }{2} = \theta (x_2^+) \;, 
\end{align} 
with the condition $x_2^+ >0$, and the inverse transform of $q^{2\nu}$ as
\begin{align}
 \frac{M_R^{1+\nu} }{\pi 2^{1-\nu} i^{1+\nu} \Gamma (-\nu)} 
\frac{\theta(x_2^+ - x_1^+) }{(x_2^+ - x_1^+ )^{2+\nu}} \exp \left( {i M_R \frac{(\vec y_2 - \vec y_1)^2 }{2 (x_2^+ - x_1^+)}} \right) \;,
\end{align}
where we used $(\vec y_2 - \vec y_1)^2 >0$ and $ x_2^+ - x_1^+ >0$. 
Then the coordinate space correlation function is 
\begin{align}
 &-2 \left(\frac{ x_2^+}{ x_1^+}\right)^{i \frac{\alpha M_R}{2}} \int \frac{d \omega}{2 \pi} \frac{d^2 k}{(2\pi )^2} 
 e^{-i \vec{k} \cdot (\vec{y}_{1}-\vec{y}_{2})} e^{ i \omega (x^{+}_{1} - x_2^+) } \cdot  
 \int \frac{d \omega'}{2 \pi} e^{- i (\omega' -\omega) x_2^+ } {\cal G}  (\omega' - \omega)  \cdot 
  {\cal F}(u_{B},\omega,\vec{k})  \nonumber \\
&=    \frac{\Gamma (1-\nu)}{ \Gamma (\nu) \Gamma (-\nu)}  
\frac{L^3 M_R^{1+\nu}  }{\pi 2^{\nu-1} i^{1+\nu} u_B^{4-2\nu}} 
\cdot \frac{\theta (x_2^+) \theta(x_2^+ - x_1^+)  }{ (x_2^+ - x_1^+)^{2+\nu}}  
 \left(\frac{ x_2^+}{ x_1^+}\right)^{i \frac{\alpha M_R}{2}}
\exp \left( {i  \frac{M_R (\vec y_2 - \vec y_1)^2 }{2 ( x_2^+ - x_1^+)}}\right) 
 \;. \label{ZeroTCorrelatorReal}
\end{align}
This is our main result for the zero temperature correlation function with $M= M_R$. We would like to pause to comment:
\begin{itemize}
\item We find that the time dependent polynomial part is the same as the result \cite{Jottar:2010vp} when $ M = M_R$. 
This is associated with the fact the the scaling dimension of a boundary scalar operator is opposite to that of the source 
of the corresponding scalar field. 
\begin{align} 
&\langle {\mathcal O}^* (x_{2}^{+} , \vec y_2) {\mathcal O}(x_{1}^{+},\vec y_1) \rangle_{\text{Aging}} 
= \left( \frac{x^{+}_{1}}{x^{+}_{2}} \right)^{-\frac{ i \alpha M_R}{2}}
 \langle {\mathcal O}^* (x_{2}^{+} , \vec y_2) {\mathcal O}(x_{1}^{+},\vec y_1) \rangle_{\text{Schr\"odinger}} \;. 
 \label{AgingSchrCorrReal}
\end{align}

\item It turns out that there exist only oscillating behaviors for $M= M_R$. Thus the same is true for real value for $\alpha M$, 
because there exists only the combination $\alpha M$. 

\item For the massless case with $\nu = 2$, we have 
\begin{align}
	{\cal F}(u_B,\omega,\vec{k}) = - \frac{L^3}{8} q^{4} \log (q^{2} ) + \cdots \;.
\end{align}
Thus the coordinate space correlation function is given by
\begin{align}
&\langle {\mathcal O}^* (x_{2}^{+} , \vec y_2) {\mathcal O}(x_{1}^{+},\vec y_1) \rangle  \nonumber \\ 
&=-
\frac{ i L^3 M_R^{3}  }{\pi } 
\cdot \frac{\theta (x_2^+) \theta(x_2^+ - x_1^+)  }{ (x_2^+ - x_1^+)^{2+\nu}}  
 \left(\frac{ x_2^+}{ x_1^+}\right)^{i \frac{\alpha M_R}{2}}
\exp \left( {i  \frac{M_R (\vec y_2 - \vec y_1)^2 }{2 ( x_2^+ - x_1^+)}}\right) 
 \;. 
\end{align}

\end{itemize}

\subsubsection{$M=i M_I$}

Let us evaluate the time integral first. 
We perform the Fourier transform for the time range $0 \leq x^+ < \infty$, and define 
$ {\cal G} (w) =  \int dx^+ e^{i w x^+} \theta (x^+) \cdot  (\alpha x^+)^{-\alpha M_I}$. Then  
\begin{align}
{\cal G} (w)
=\alpha^{-\alpha M_I} |w|^{-1+\alpha M_I} \Gamma (1-\alpha M_I)  \left(- i \cos (\frac{\pi \alpha M_I}{2}) \text{sign} (w)  + \sin (\frac{\pi \alpha M_I}{2})  \right) \;. 
\label{GMI}
\end{align}  
where $w=\omega' -\omega$.
Note that these results are essentially the same as those evaluated by Laplace transform in, so called, $s$-domain. 

Thus we get the momentum correlation function as 
\begin{align}
	&\langle {\cal O}^* (\omega',\vec{k}') {\cal O}(\omega, \vec{k}) \rangle 
	= -2 (2\pi)^{-3}  \delta (\vec{k'} - \vec{k} )~ {\cal G} (\omega' -\omega) ~ {\cal F}(u_{B},\omega,\vec{k}) \;. 
\end{align}  
Again, ${\cal F}$ is given in equation (\ref{FFfunction}), and $q=\sqrt{\vec k^2 + 2M \omega}$. 
Here we can have $\vec k=0$ due to the spatial translational invariance, but not for $\omega$ 
because time translation symmetry is broken. 

Let us calculate the coordinate space correlation function as  
\begin{align} 
&\langle {\mathcal O}^* (x_{2}^{+},\vec{y}_{2})  {\mathcal O}(x_{1}^{+},\vec{y}_{1}) \rangle \\
&\quad = \int \frac{d \omega'}{2 \pi} \frac{d^2 k'}{(2\pi )^2} \frac{d \omega}{2 \pi} \frac{d^2 k}{(2\pi )^2} 
e^{i \vec{k'} \cdot \vec{y}_{2} -i \vec k \cdot \vec{y}_{1} } e^{ -i \omega' \cdot x^{+}_{2} + i \omega \cdot x_1^+} 
\left( \alpha^2 x_1^+ x_2^+ \right)^{\frac{\alpha M_I}{2}}
\langle {\cal O}^* (\omega',\vec{k}') {\cal O}(\omega, \vec{k}) \rangle \;. \nonumber 
\end{align} 
After carrying out the integral for the $\omega' $ first 
\begin{align}
  \int \frac{d \omega'}{2 \pi} ~e^{  i ( \omega' - \omega)  x^{+}_{2}} ~ {\cal G} (\omega' -\omega) 
 = \theta (x_2^+)~\alpha^{-\alpha M_I}  |x_2^+|^{- \alpha M_I} \;, 
\end{align}
the expression reduces to the previous case for the time independent part, whose calculation is done similarly. We get 
\begin{align}
&\langle {\mathcal O}^* (x_{2}^{+},\vec{y}_{2})  {\mathcal O}(x_{1}^{+},\vec{y}_{1}) \rangle \nonumber  \\ 
&=  
-2 \theta (x_2^+)~\left( \alpha^2 x_1^+ x_2^+ \right)^{\frac{\alpha M_I}{2}} ~\alpha^{-\alpha M_I}  |x_2^+|^{- \alpha M_I}  \int  \frac{d \omega}{2 \pi} \frac{d^2 k}{(2\pi )^2} 
 e^{-i \vec{k} \cdot (\vec{y}_{1}-\vec{y}_{2})} e^{ i \omega (x^{+}_{1} - x_2^+) }  {\cal F}(u_{B},\omega,\vec{k})  \nonumber \\
 &=    \frac{\Gamma (1-\nu)}{ \Gamma (\nu) \Gamma (-\nu)}  
\frac{L^3 M_I^{1+\nu}  }{\pi 2^{\nu-1}  u_B^{4-2\nu}} 
\cdot \frac{\theta (x_2^+) \theta(x_2^+ - x_1^+) }{ (x_2^+ - x_1^+)^{2+\nu}}   
\cdot \left( \frac{x_2^+}{x_1^+} \right)^{-\frac{\alpha M_I}{2} }   \cdot \exp \left( -{ \frac{M_I (\vec y_2 - \vec y_1)^2 }{2 ( x_2^+ - x_1^+)}}\right)  \;. 
\label{ZeroTCorrelatorIm}
\end{align}
This is the result we are looking for. This confirms that the correlation function has dissipative behavior for $M= i M_I$. 
Thus 
\begin{itemize}
\item We find that the time dependent polynomial part is identical to the result \cite{Jottar:2010vp} when $ M = M_R$. 
This is associated with the fact the the scaling dimension of a boundary scalar operator is opposite to that of the source 
of the corresponding scalar field. 
\begin{align} 
&\langle {\mathcal O}^* (x_{2}^{+} , \vec y_2) {\mathcal O}(x_{1}^{+},\vec y_1) \rangle_{\text{Aging}} 
= \left( \frac{x^{+}_{2}}{x^{+}_{1}} \right)^{-\frac{  \alpha M_I}{2}}
 \langle {\mathcal O}^* (x_{2}^{+} , \vec y_2) {\mathcal O}(x_{1}^{+},\vec y_1) \rangle_{\text{Schr\"odinger}} \;. 
 \label{AgingSchrCorrIm}
\end{align}

\item There exist only the combination $\alpha M$, and thus the result is true for imaginary $\alpha M$.

\item For the massless case with $\nu = 2$, we have the coordinate space correlation function similar to the $M=M_R$ case as
\begin{align}
&\langle {\mathcal O}^* (x_{2}^{+} , \vec y_2) {\mathcal O}(x_{1}^{+},\vec y_1) \rangle  
=-
\frac{ i L^3 M_I^{3}  }{\pi } 
\cdot \frac{\theta (x_2^+) \theta(x_2^+ - x_1^+)  }{ (x_2^+ - x_1^+)^{2+\nu}}  
 \left(\frac{ x_2^+}{ x_1^+}\right)^{-\frac{\alpha M_I}{2}}
\exp \left( {-  \frac{M_I (\vec y_2 - \vec y_1)^2 }{2 ( x_2^+ - x_1^+)}}\right) 
 \;. 
\end{align}

\end{itemize}

\begin{figure}[!ht]
\begin{center}
\begin{tabular}{cc}
	 \includegraphics[width=0.6\textwidth]{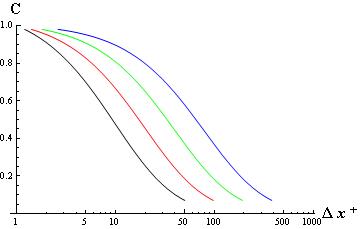}
 \end{tabular}
 \caption{
 Plot of the equation (\ref{TwoTimeCorr}) with $\nu = 0,~ \alpha M_I = 1, ~\alpha M_R = 0$.
 The horizontal and vertical axes are $\Delta x^+ =x_2^+ - x_1^+$ and two-time correlation function 
 $C= C(x^{+}_{2}, x^{+}_{1})$,
 respectively.  The four plots, black, red, green and blue (from left to right)  are for the particular 
 waiting time $x_1^+ = 25, 50, 100$, and $200$, respectively. 
 They show that the older is the system the slower it relaxes. 
 This plot is to be compared with that of the numerical results in \cite{Helkel:2007nept}.}
\label{fig:AgingPlot}
\end{center}
\end{figure}

\subsection{Two-time Correlators}

One of the important physical observables considered in the literature (see {\it e.g.} 
\cite{Cugliandolo:2002dy}\cite{Henkel:2002vd}) is two-time correlation functions 
at the same space positions, which we refer to $C(x^{+}_{2}, x^{+}_{1})$ 
\begin{align}
C(x^{+}_{2}, x^{+}_{1}) &= \langle {\mathcal O}^* (x_{2}^{+} , \vec y_2) 
{\mathcal O}(x_{1}^{+},\vec y_1) \rangle _{\vec y_1 = \vec y_2}  
\equiv (x_1^+ )^{-\lambda_b} ~f_C \left( \frac{x_2^+}{x_1^+} \right)  \;, 
\label{TwoTimeCorr}
\end{align}
where $
f_C \left( x \right) = x^{-\lambda_C /{\rm z}}$,
and z is the dynamical exponent. 
$x_1^+$ and $x_2^+$ are called a waiting time and a response time, respectively. 
This scaling behavior is expected to apply for the "aging regime": 
$ x_1^+, x_2^+ \gg x_{mi}^+ $ and $ x_2^+ - x_1^+ \gg x_{mi}^+$, where
$x_{mi}^+$ is a microscopic time scale of a given system. 

Our dual field theory possesses aging invariance, inherited from the bulk gravity and hence exhibits two 
important characteristic features of aging system, namely the time dependence and the existence of dynamical scaling.  
Furthermore, for $M_R=0$ and for $x_2^+ \gg x_1^+ $ , which is physically interesting aging regime, 
our two-time correlation function decays slowly following power law, like (\ref{TwoTimeCorr}), with
\begin{align}
\lambda_b = \nu + 2 , \qquad \lambda_C 
=  {\rm z} \left(\nu + 2 +\frac{ \alpha M_I}{2}\right) \;,
\label{exponents}
\end{align}
where ${\rm z}=2$ for our case. 
These two parameters $\lambda_b$ and $\lambda_C$ are our physical exponents associated 
with the two-time correlation function. 
The  dissipating behaviors of the correlation function are plotted in figure \ref{fig:AgingPlot} with several different waiting time.  
We clearly see the typical aging behavior: older systems relax in a slower 
manner than younger systems, and thus the correlation between two different time 
$x_1^+$ and $x_2^+$ has more correlation for the bigger waiting time $x_1^+$ in our 
correlation function $C(x^{+}_{2}, x^{+}_{1})$.

\subsection{Comments on Schr\"odinger Background}

In this section, we would like to briefly comment on the aging construction on the Schr\"odinger 
background \cite{Jottar:2010vp} with our approach, especially with the complex $M$ and without 
complexifying the spacetime geometry.
 
Schr\"odinger background can be embedded in string theory 
in a nice way using null Melvin twist \cite{Gimon:2003xk}\cite{Alishahiha:2003ru}.
Previous sections already confirmed that the results of the scalar wave solution 
and the two-point correlation function are the same for both cases even though these two approaches 
are rather different. This is mainly because the Schr\"odinger symmetry is large enough to 
restrict the physical properties of two point correlation functions. It is still worthwhile to revisit the aging construction for 
the Schr\"odinger background with our approach. 

Let us consider the following metric with the change $x^+ \rightarrow t$, $u\rightarrow z$ 
and $x^- \rightarrow \xi$  
\begin{align} 
ds^{2} &= \frac{L^{2}}{z^2}\left( dz^2  - \frac{1}{z^{2}} \,dt^{2} -2dtd\xi  + d\vec{y}^{2}\right) 
+ \frac{L^{2}}{z^2}\left( \frac{2\alpha}{z}\,dzdt -   \frac{\alpha  }{  t} \,dt^{2} \right) \;,
\end{align}
where extra matter contents, which are not explicitly written here, are necessary to support this background. 
Following the previous section, we couple a scalar field to this background. 
We find the identical radial differential equation as the one in AdS in light-cone if we change, 
$m^2  \rightarrow m'^2 = m^2 + M^2 / L^2$ \cite{Goldberger:2008vg}. 

For the complex parameter $M=M_R + i M_I$, 
the form of the corresponding solution is the same as (\ref{AgingWaveFunction})  
with $q =\sqrt{\vec k^2 +2 M \omega}$ and $\nu'= \sqrt{4+L^2 m^2 +M^2 }$. 
These parameters were used in the Schr\"odinger background \cite{Son:2008ye}\cite{Balasubramanian:2008dm}. 
Again we confirm that the aging scalar wave solution is given by the scale-invariant factor, 
$\left(  \alpha x^+ / u^2 \right)^{\frac{i\alpha M}{2}}$, times the Schr\"odinger wave solution as 
advertised in \cite{Jottar:2010vp}. This is given in (\ref{FinalWaveFunction}). 
With this in hand, it is straightforward to evaluate the two-point correlation function, which is also the 
same as that of the AdS in light-cone and is given in equations 
(\ref{ZeroTCorrelatorReal}) for $M=M_R$ and (\ref{ZeroTCorrelatorIm}) for $M=i M_I$.
Thus, at zero temperature, the physical properties of Aging in light-cone are the same as those of the Aging background 
for the operators with the same scaling dimensions, $\nu = \nu'$.
The scaling dimension of the operator $\phi$ is complex in general for complex $M$. 
If we consider either real $M=M_R$ or pure imaginary $M= i M_I$, which are physically interesting cases, 
the scaling dimension is still real. It is interesting to notice that 
the scaling dimension $\Delta_\phi = 2-\nu' $ 
decreases for the real $M=M_R$, while it increases for imaginary $M=i M_I$.  

All the other discussions and physical properties remain unchanged compared to the Aging in light-cone at zero temperature. 
Interestingly enough,  as we will see later, these do not persist at the finite temperature.

\subsection{Comments on Dual Field Theory} 

In the context of the holographic condensed matter application, the field theory dual of the gravity 
description is not explicitly known in most cases. For the case of the Schr\"odinger 
geometry, the situation is much better because the construction are explicitly known as 
Discrete Light-Cone Quantization (DLCQ) for AdS in light-cone and null Melvin twist 
for the Schr\"odinger background. 
The basic field theory dual of AdS in light-cone is DLCQ of the ${\cal N}=4$ Super Yang-Mills theory 
proposed in \cite{Ganor:1997jx}\cite{Kapustin:1998pb} as explained in \cite{Maldacena:2008wh}. 
It is further required to project the conserved particle number $M$ to a single sector, which  
is related to the isometry of the gravity backgrounds and has distinguished role 
\cite{Son:2008ye,Goldberger:2008vg,Barbon:2008bg,Kim:2010tf}. 
We are required to focus on a single sector of the particle number.
For further developments along the Schr\"odinger background with null Melvin twist, 
see \cite{Adams:2008wt}. In this section we would like to comment on some properties 
directly related to the time dependent generalizations of the setup, which apply for both 
AdS in light-cone and Schr\"odinger backgrounds.  

The differential equation (\ref{BlukScalarEq}), we consider at the end of section \ref{sec:AgingLC}, 
has an explicit time dependence. Thus the dual boundary field theory is expected to 
have not Schr\"odinger but aging symmetries. From the observation that the equation 
(\ref{BlukScalarEq}) factorize 
into $u-$dependent and $u-$independent parts, we factorize $\phi = \varphi(x^+, \vec{y}) ~f(u)$ and 
find that the boundary wave function $\varphi$ satisfies the following differential 
equation
\begin{align}
2M \left(i  \frac{\partial}{\partial x^+}  +  \frac{\alpha  M}{2 x^+}   \right) \varphi 
+ \vec \nabla ^2 \varphi 
-v^2 \varphi =0
  \;,
\label{BoundaryScalarEq}
\end{align}
where $v^2$ is the eigenvalue of the radial differential equation. The term  $v^2\varphi$ can be interpreted as the one coming from the variation of the Landau-Ginzburg potential, i.e. $\frac{\delta {\cal V}_{LG}}{\delta \varphi}=v^2\varphi$.   One may generate the general Landau-Ginzburg potential by considering the self-interacting scalar fields in bulk. 
It is clear that our  geometric realization of aging symmetry actually constrains
the nature of the time dependence of the equation of the dual field theory as the form 
\begin{align}
\frac{\alpha  M^2}{ x^+}  \varphi \;.
\end{align}
It is interesting to note that if we can add noise contributions the differential equation 
becomes the Langevin equation \cite{HJK1}. 

It is not difficult to solve the differential equation (\ref{BoundaryScalarEq}) and 
read off the time dependent part of the solution. Compared to the $\alpha=0$ case we get
\begin{align}
\varphi \quad \longrightarrow  \quad  (x^+)^{\frac{i\alpha M}{2}} ~\varphi \;.
\end{align}
Thus with aging symmetry, the boundary scalar field will acquire the 
explicit time-dependence of this form. Note that we can also infer the same conclusion from the 
local transformation (\ref{CoordinateChange}) which contains essential effects to 
the bulk side of the story. Furthermore, the parameter $v$ contains other effects 
due to the renormalization of the radial coordinate. 
It turns out that, for real $M$, there is no net effects of this radial 
wave function renormalization on physical observables such as the two-point correlation function
as we see in (\ref{FinalWaveFunction}) and (\ref{ZeroTCorrelatorReal})(\ref{ZeroTCorrelatorIm}).

\section{Finite Temperature Aging Holography}

We start with the Einstein-Hilbert action with cosmological constant, which admits the planar black hole solution 
\begin{eqnarray}
ds^{2} &=& \left( \frac{r}{L} \right) ^{2} \left[ \frac{1-h }{4b^{2}} (dx^{+})^{2}-(1+h )dx^{+}dx^{-} + (1-h )b^{2} (dx^{-})^{2} + d\vec{y}^{2} \right]\nonumber \\
 &+&\left( \frac{L}{r} \right) ^{2} \frac{1}{h } dr^{2} \;, \qquad \qquad  h = 1 - \frac{r_H^4}{r^4}\;.
\end{eqnarray}

We would like to exploit the idea that the aging metric is locally Schr\"odinger, and thus we can
get the Aging black hole from the Schr\"odinger black hole using the coordinate transformation as
\begin{eqnarray}
x^{-} &\rightarrow& x^{-} +\frac{\alpha}{2} \textrm{ln} \left( r^{2} x^{+} \right) \;. 
\label{SingularTR}
\end{eqnarray}
The black hole metric in the simplest form can be written as
\begin{align}
  ds^2 =& \bigg(\frac{r}{L}\bigg)^2 \bigg\{ - \frac{h}{1-h}b^{-2} dx^{+2}
  + (1-h)b^2 \bigg( dx^- +\frac{\alpha}{r}d r -\left(- \frac{\alpha}{2x^{+}} + \frac{1}{2b^2}\frac{1+h}{1-h} \right)dx^+ \bigg)^2
  + d \vec y^2  \bigg\}   \nonumber\\
  &+ \bigg(\frac{L}{r}\bigg)^2 h^{-1} dr^2   \;.
\end{align}
We would like to consider this finite temperature metric to construct the correlation functions.

\subsection{Equation of Motion} \label{sec:finiteTeq}

We consider a probe scalar on the Aging black hole background and try to 
evaluate the two-point correlator similar to the previous section. The equation of motion is given by 
\begin{align}
&\frac{1}{\sqrt{-g}} \partial_{\mu} \left( \sqrt{-g} g^{\mu\nu} \partial_{\nu} \phi \right) -m^{2} \phi =0 \;. 
\label{probeEq}
\end{align}
The metric depends only on $(x^{+},r)$ and the equations of motion can be expressed as 
\begin{align}
\frac{r}{L^{2}} (4+h)\partial_{r}\phi - \frac{4\alpha}{L^{2}} \partial_{-} \phi - \frac{L^{2}}{r^{2}} \frac{\alpha b^2}{2(x^{+})^{2}}\frac{1-h}{h} \partial_{-} \phi +g^{\mu\nu} \partial_{\mu}\partial_{\nu} \phi -m^{2} \phi=0 \;.
\end{align}
 As was done in the previous section, we replace the derivative of the $x^-$ coordinate as $\partial_{x^-} = iM$. The mode expansion becomes of the form, 
\begin{align}
\phi(r,x^{+},\vec{y}) = \int \frac{d\omega}{2\pi} \frac{d^{2}k}{(2\pi)^{2}} \exp^{-i\vec{k}\cdot \vec{y}} \phi_k ( r,x^{+})  \phi_{0}(\omega,\vec{k})\;,   
\end{align}
where $\phi_k ( r,x^{+})= T_{\omega}(x^{+}) f_{\omega,\vec{k}} (r)$. 

The equation of motion in $r$ coordinate is complicated. It is slightly better to go to a different 
coordinate using $z= \frac{r_H^2}{r^2}$.
Then the differential equation is 
\begin{align}
& \frac{L^4}{4 z h r_H^2} \left( \frac{1+h}{h} M \frac{\hat{D} T_{\omega}}{T_{\omega}}-\frac{(1-h)b^{2}}{h} \frac{\hat{D}^{2} T_{\omega}}{T_{\omega}} \right) \nonumber  \\
&  =
\frac{f_{\omega,\vec{k}}''}{ f_{\omega,\vec{k}}} 
- \frac{\left( -i \alpha M h-h +2 \right)}{z h}   \frac{f_{\omega,\vec{k}}'}{f_{\omega,\vec{k}}}   
-\left( \frac{4i\alpha M + \alpha^2 M^2 h + m^2 L^2 }{4 z^2 h}+ \frac{\vec k^{2}  
- \frac{M^{2}}{4b^{2}}\frac{1- h }{ h }}{4 z h r_H^2 /L^4} \right)  \;, 
\label{FiniteTDiff}
\end{align}
where $f'=\partial_z f$, and we used 
\begin{align}  
\hat{D} \equiv i \frac{\partial}{\partial {x^+}} +\frac{\alpha M}{2x^{+}} \;, \qquad h = 1-z^2 \;.
\label{TimeOp}
\end{align}
One qualitatively different feature from zero temperature case  is that this finite temperature version of 
the differential equation does have
a second order time derivative term. Yet as we see in the first line of equation (\ref{FiniteTDiff}), 
this does not generate different dynamics along time direction compared to the zero temperature case 
because the time dependent parts of the differential equation, equation (\ref{FiniteTDiff}) factorize in a special way.

The time dependent part can be solved as follows: 
\begin{eqnarray}
 \left( \frac{\alpha  M}{2 x^+} +  i  \partial_{+} \right) T_{\omega}   = \omega T_{\omega} \quad \longrightarrow \quad
T_{\omega}(x^+) =  c_1  \exp^{-i \omega x^+} (x^+)^{\frac{i\alpha M}{2}} \;.
\end{eqnarray}
Thus, we would like to stress that, the time dependent part of the scalar wave solution at finite temperature is exactly the same as that of the zero temperature case. Anticipating the appearance of the scaling invariant combination,  $x^+/z$,  we factorize  
as $f_{\omega,\vec{k}} (z) = z^{-\frac{i\alpha M}{2}} \tilde g_{\omega,\vec{k}} (z)$
and obtain the radial equation:
\begin{align}
&\tilde g ''- \frac{1+ z^2 }{z (1-z^2) } \tilde g ' - \frac{  m^2 L^2 }{4 z^2 (1-z^2)} \tilde g 
+   \frac{{\bf w}^2 z }{ (1-z^2)^2} \tilde g
-  \frac{{\bf q}^2 }{z (1-z^2)} \tilde g = 0 \;, \nonumber \\
&\text{where} \qquad {\bf w} =  \frac{M/(2b^2)- \omega }{2 \pi  T} \;, \qquad {\bf q}^2 =  \frac{2 M \omega + \vec k^2}{(2 \pi b T)^2}\;,
\label{MassiveEq}
\end{align}
where used $r_H =\pi L^2 b T$. 
The equation includes the term which has  the Schr\"odinger invariant combination 
$2 M \omega + \vec k^2$, explicitly, and the term which is partially broken due to the other combination 
${\bf w} \sim M/(2b)-b \omega$ at finite temperature. 
The term proportional to ${\bf w}$ is dominant at the horizon 
$z=1$ and important for the calculation of the correlation functions. 
Interestingly enough, the combination of chemical potential $\frac{1}{2b^2}$ and conserved charge $M$ of 
$x^-$ direction of the time-independent Schr\"odinger black hole 
shows up with the energy of the probe scalar $\omega$. 
In what follows, we consider the low energy regime in which  
the magnitude of the combination $\frac{M}{2b^2}$ is bigger than $\omega$, which  corresponds to ${\bf w} >0$.%

Note that this equation is nothing but the radial equation for the AdS planar black hole in light-cone. 
Thus we explicitly check that the radial part of the time independent aging wave solution is the same 
as that of the AdS in light-cone case. This has an important implication that the time independent part 
of the aging wave solution and correlation functions are those of the AdS in light-cone at least in 
momentum space with some modification, which can be identified and thus isolated. 
     
The full solution is 
\begin{eqnarray}
\phi(z,x^+, \vec y) = \int \frac{d^2 k}{(2\pi )^2}\frac{d \omega}{2 \pi} e^{i \vec{k} \cdot \vec{y} -i \omega x^+}   
\left( \frac{\alpha x^+}{z} \right)^{\frac{i\alpha M}{2}} ~c~ \tilde g_{\omega,\vec{k}} (z) ~ \phi_0 (\omega,\vec{k}) \;, 
\label{FinTsol}
\end{eqnarray}
where the solution has the desired scaling form with an appropriate normalization for the correlator calculation.
We concentrate on the momentum space correlation functions from now on.

\subsection{Finite Temperature Correlation Functions} 

In this section, we would like to evaluate the scalar correlation function of the Schr\"odinger 
geometry via AdS in light-cone with finite parameters $M$, $\omega$ and $\vec k$, which is fairly 
non-trivial even in this simple setup. First, we consider the time dependence of the coordinate space correlation 
functions and outline the general properties of the correlation functions at finite temperature. 
After that we concentrate on the momentum space correlation functions starting with analytic approaches in general. 
And then we evaluate the correlation functions analytically for some special cases and obtain the expression for 
the shear viscosity, which is presented in appendix \ref{sec:viscosity}. 
Finally we get the full features with numerical studies. 
From the numerical study, we observe two distinct peaks in momentum space correlation functions, 
one broad peak at the $(\omega, k)=(0,0)$ and another at finite values of $\omega$ and $k$, 
which are explained in detail at the end of this section.

\subsubsection{General Features} 

In section (\ref{sec:finiteTeq}), we have some important observations regarding the time dependence of the 
scalar equation of motion. Combining with the experience from the zero temperature case in section 
(\ref{ZeroTCorrelationFunction}), we would like to 
see the general time dependence of the coordinate space correlation functions.  

Following the steps done in section (\ref{ZeroTCorrelationFunction}), we rewrite the solution (\ref{FinTsol}) 
as 
\begin{eqnarray}
\phi(z,x^+, \vec y) = \int \frac{d^2 k}{(2\pi )^2}\frac{d \omega}{2 \pi} e^{i \vec{k} \cdot \vec{y} -i \omega x^+}   
\left( \alpha x^+ \right)^{\frac{i\alpha M}{2}} ~ f_{\omega,\vec{k}} (z) ~ \phi_0 (\omega,\vec{k}) \;, 
\label{FinTsolTF}
\end{eqnarray}
where $f_{\omega,\vec{k}} (z) = c z^{-\frac{i\alpha M}{2}} \tilde g_{\omega,\vec{k}} (z)$. 
Then the on-shell action has the form 
\begin{align}
	S[\phi_0] 
	&= \int d^3 x  \sqrt{-g} ~\phi^* (z,x^+, \vec y) ~\left(g^{zz} \partial_z + i M g^{z-}  \right) \phi (z,x^+, \vec y) \big |_{z_B} \;. 
\end{align}
This can be recast using the equation (\ref{FinTsolTF}) as 
\begin{align}	
	&\int d x^+  ~\theta (x^+) ~\frac{d \omega'}{2 \pi} \frac{d \omega}{2 \pi} e^{-i (\omega' -\omega) x^+} 
	\left( \alpha x^+ \right)^{-\frac{i\alpha (M^*-M)}{2}} 
	\nonumber \\
	&\qquad \times 
	\int d^2 y  \int \frac{d^2 k'}{(2\pi)^2} \int \frac{d^2 k}{(2\pi)^2} 
	e^{i(\vec k' - \vec k) \cdot \vec y} ~ \phi_0^* (\omega',\vec{k'}) {\cal F}(u, \omega', \omega, \vec{k'}, \vec{k}) \phi_0
	(\omega,\vec{k}) \big |_{u_B} \;, 
	\label{wholeEQ2}
\end{align}
where $\theta (x^+)$ represents the physical region as $0 \leq x^+ < \infty$, and ${\cal F}$ is given by 
\begin{align}
	{\cal F} (z, \omega', \omega, \vec{k'}, \vec{k})
	&= \sqrt{-g} f_{\omega',\vec{k'}}^* (\omega',\vec{k'},z) \left( g^{zz} \partial_z
	+ i M g^{z-} \right) f_{\omega,\vec{k}} (\omega,\vec{k},z) .
\end{align}
Note that the spatial integration along $\vec y$ can be done to give delta function $\delta^2 (\vec k' -\vec k)$. 
One can bring the $z^{\pm i \frac{\alpha M}{2}}$ factors in $f$ and $f^*$ together to cancel each other, which
removes the second part in $ {\cal F}$. Then time independent radial part ${\cal F}$ of momentum correlation function is given by
\begin{align}
{\cal F} (z, \omega', \omega, \vec{k'}, \vec{k})
	&=
\frac{b^4 \pi ^4 L^{5} T^4}{2 z^3} \left(  \frac{\tilde g_{\omega',\vec{k'}} (z)}{\tilde g_{\omega',\vec{k'}} (z_B)}  \right)^*  \frac{4 z^2 h }{L^2} \partial_z \left(  \frac{\tilde g_{\omega,\vec{k}} (z)}{\tilde g_{\omega,\vec{k}} (z_B)}  \right) \;. 
\label{TFFfunction}
\end{align}
It turns out that evaluating this correlation function is highly nontrivial because the differential 
equation is not analytically tractable for the AdS in light-cone except some special limits, while it 
is even more complicated for the Schr\"odinger background. 

Before explaining the time dependent part of the correlation function, we would like to comment on our choice of the incoming 
boundary condition for the equation (\ref{MassiveEq}), which is not clear due to the 
contribution of $M$ in the ${\bf w}$. We choose our  incoming boundary condition as
\begin{align}
e^{ -i \omega x^+} (1-z)^{i {\bf w}/2} \propto
~e^{ -i \omega \left( x^+ + z_* \right)}
\end{align}
where $z_* = \frac{\ln (1-z)}{4\pi T}$. We come back to this radial dependent part of the correlation function 
with several different approaches below.

As discussed in the section \ref{sec:ZeroTemp}, this is not the end of the story. 
The time dependent parts turn out to be the same as those of the zero temperature case because 
their time dependences in the wave solution (\ref{FinTsol}) are the same. 
This is explained below equation (\ref{TimeOp}). Thus the time dependent parts of the correlation 
function can be explicitly computed and are given by ${\cal G}$ in equations (\ref{GMR}) and (\ref{GMI}). 
Explicitly, we get  
\begin{align}
	&\langle {\cal O}^* (\omega',\vec{k}') {\cal O}(\omega, \vec{k}) \rangle
	= -2 (2\pi)^{-3}  \delta (\vec{k}' - \vec{k} ) ~ {\cal F}(u_{B},\omega,\vec{k})~ {\cal G}  (\omega' - \omega) 
	\;,
\end{align}
where ${\cal F}$ is given in equations (\ref{TFFfunction}). By performing the $\omega'$ integration, 
the coordinate space correlation function is give by 
\begin{align} 
&\langle {\mathcal O}^* (x_{2}^{+},\vec{y}_{2})  {\mathcal O}(x_{1}^{+},\vec{y}_{1}) \rangle 
=   \theta (x_2^+)~\left( \frac{x_2^+}{x_1^+} \right)^{-i \frac{\alpha M}{2} }
 \langle {\mathcal O}^* (x_{2}^{+},\vec{y}_{2})  {\mathcal O}(x_{1}^{+},\vec{y}_{1}) \rangle _{\text{Schr\"odinger}} \;.
 \end{align}
Where $M$ can be either $M=M_R$ or $M=i M_I$. The latter case is related to the dissipative case.   

It is not straight forward to evaluate the general momentum space correlation functions analytically, not to mention the 
coordinate space correlation functions. Thus we would like to concentrate on the momentum space correlation function 
for the rest of the section.

\subsubsection{Analytic Approaches} 

Let us outline the analytic approach we considered and summarize our observations.

The differential equation (\ref{MassiveEq}) has 4 regular singular points, which, in general, 
can not be solved with systematic approach. If we take 
\begin{align}
\tilde{g}(z)=z^{\frac{1+\bar \gamma}{2}}(z-1)^{\frac{-1+\bar \delta}{2}}(z+1)^{\frac{-1+\bar \epsilon}{2}} G(z) \;.
\end{align} 
The differential equation (\ref{MassiveEq}) can be cast into the Heun's equation 
\begin{align}
& G''(z)+\left(\frac{\bar \gamma}{z}+\frac{\bar \delta}{z-1}+\frac{\bar \epsilon}{z+1}\right) G'(z) 
+\frac{\bar \alpha \bar  \beta z -\bar q}{z(z-1)(z+1)} G(z)=0 \;, \\
\text{where} \qquad & 4 \bar \alpha \bar  \beta =  \nu^{2} -2 + 2(\bar \gamma \bar \delta +\bar \delta \bar \epsilon + \bar \epsilon \bar \gamma) \;, \quad 
-2\bar q=  \bf{q}^{2} +\bar \gamma \bar \delta + \bar \gamma \bar \epsilon \;, 
\end{align}
with the special values of 
\begin{align}
\bar \gamma=1\pm \nu \;, \qquad \bar \delta=1\pm i {\bf w} \;, \qquad 
\bar \epsilon=1\pm \bf{w} \;.
\end{align}
The general solutions of  this equation are known as the Heun functions 
$
Hl(-1, \bar q ; \bar \alpha, \bar \beta, \bar \gamma, \bar \delta ; z) \;
$.
In general, we have 8 different combination of the parameters ${\bar\gamma, \bar\delta, \bar\epsilon}$, 
out of which only 4 solutions are physical 
due to the incoming boundary condition we would like to impose at $z=1$. 
That corresponds to the choice $\bar \delta=1 + i \bf{w}$ due to our choice of ${\bf w}$ parameter given in 
equation (\ref{MassiveEq}). For special values of parameters, the Heun functions might provide 
analytic solutions.  

We observe that our differential equation (\ref{MassiveEq}) is similar to the equation 
(3.4) of \cite{Policastro:2001yb}, even though they have very different setup. 
We can rewrite the differential equation (\ref{MassiveEq}) in a slightly different way 
to make the comparison straightforward 
\begin{align}
\tilde {\tilde g} '' -\frac{1+z^2 }{z(1-z^2) }  \tilde {\tilde g} '  
-\left( \frac{ L^2 m^2  }{4z^2 (1-z^2)} +\frac{  Q^2 - z^2 Y^2}{z(1-z^2)^2}  \right) \tilde {\tilde g }  =0 \;.
\end{align}
where $Y^2 =  \frac{\vec k^2 +( M/(2b)+b \omega)^2}{4\pi^2 b^2 T^2}$, 
$Q^2 = \frac{2 M \omega + \vec k^2}{4\pi^2 b^2 T^2}$.
Note that, in general, $Y^2 \geq Q^2$, where the equality holds when $M =  2 \omega b^2$. 
One of the crucial difference between this equation and (3.4) of \cite{Policastro:2001yb} lies in the 
last term. In our case, $Y^2$ has always larger contribution than $Q^2$, and the term 
proportional to $Y^2$ is multiplies by $z^2$. On the other hand, the corresponding 
term in (3.4) of \cite{Policastro:2001yb} can be neglected to give a constant numerator, 
which make possible to proceed further. 
Due to the difference, we can not use the approach adapted in \cite{Policastro:2001yb} in a reliable way 
except for a very special case, which we don't pursue further here. 
 
In subsequent sections, we consider high temperature limit, called hydrodynamic limit, and 
low temperature limit in some details. And then we evaluate the differential equation 
using numerical method to see the full features of the correlation functions.

\subsubsection{High Temperature Correlation Functions} \label{sec:HighTCorr}

For the high temperature limit, we have small parameters ${\bf w}$ and ${\bf q}^2$, which can be used as 
expansion parameters in equation (\ref{MassiveEq}). Following \cite{Son:2002sd}, we can derive similar 
results \cite{Herzog:2008wg}\cite{Adams:2008wt}\cite{Kim:2010tf}. For $m=0$ case, we calculate ${\cal F}$ 
using incoming boundary condition at the horizon  
\begin{align}
	{\cal F} &=  2 b^4 \pi ^4 L^{3} T^4 \left\{ - \frac{{\bf  q}^2}{z_B} +\left( i {\bf w} + {\bf  q}^2  \right)   + {\cal O}\left({\bf w}^2, 
	 {\bf w} {\bf q}^2 , {\bf q}^4\right) \right\}\;, 
	 \label{FFFunctionHighT}
\end{align}
where we used 
\begin{align} \tilde g_{\omega,\vec{k}} (z) = (1-z)^{i {\bf w}/2} \left(1 - \left(i\frac{ {\bf w}}{2} + {\bf q}^2\right) \ln \frac{1+z}{2} + \cdots \right) \;.
\end{align}
Note that the form of the first non-trivial contributions are the same as the results of \cite{Son:2002sd} 
even though the differential equations are different. Yet, the physical implication seems to be more complicated. 
In general we can not claim to have ${\bf q}=0$ for the zero momentum case $\vec k =0$ 
because of the parameter $M$. 
The first term diverges for the strict $u_B \rightarrow 0$ limit, while the second term gives  finite contributions as  
\begin{align}
\text{Re} ~{\cal F} &= 2 b^4 \pi ^4 L^{3} T^4 ~ \left( -\frac{  M_I  }{4  \pi b^2 T} + \frac{2 M_R \omega + \vec k^2}{(2 \pi b T)^2} \right) \;, \\
\text{Im} ~{\cal F} &=  2 b^4 \pi ^4 L^{3} T^4 ~ \left( \frac{M_R -2 b^2 \omega }{4  \pi b^2 T} + \frac{2 M_I \omega}{(2 \pi b T)^2} \right) \;.
\end{align}
Where we put the result with $M=M_R + i M_I$ for notational simplicity.

Let us concentrate on the case ${\bf q}=0$. This explicitly means that $ 2 M_R \omega + \vec k^2 = 0$ and 
$M_I = 0$. Thus the momentum correlator has an extra contribution in imaginary part of the correlator 
compared to relativistic case \cite{Son:2002sd}  
\begin{align}
\text{Re} ~{\cal F} =  0 \;, \qquad 
\text{Im} ~{\cal F} =  2 b^4 \pi ^4 L^{3} T^4 ~ \frac{M_R -2 b^2 \omega }{4  \pi b^2 T} \;. 
\end{align}
If one considers the case of zero spatial momenta $\vec k=0$ and thus $M_R=0$, 
we get 
\begin{align}
\text{Im} ~{\cal F} =  - b^4 \pi ^3 L^{3} T^3  \omega \;. 
\label{ImFFfinal}
\end{align}
For the application to shear viscosity, see appendix \ref{sec:viscosity}.

\subsubsection{Low Temperature Correlation Functions }  \label{sec:LowTCorr}

With a change of the function $\tilde g(z) = \sqrt{ z-\frac{1}{z} }~ g(z) $, we get the
differential equation for $g(z)$ 
\begin{align}
&g'' = \left({\bf w}^2 F  + H  \right) g  \;,   \label{DiffLowT} \\
F =  \frac{(1-z^2) {\bf s}^2 - z^2}{z \left(1-z^2\right)^2}\;, \qquad
& H = - \frac{-3+6 z^2+z^4 - m^2 L^2 (1-z^2)}{4 z^2 \left(1-z^2\right)^2} \;,
\end{align}
where, in the low temperature regime, $ {\bf q}^2, {\bf w}^2 \gg 1$ and  
${\bf s} ^2=  \frac{{\bf q}^2}{{\bf w}^2}  = \frac{ 2 M_R \omega + \vec k^2}{(M_R/(2b)-b \omega)^2}  
\approx  {\cal O} (1)$.
Then the term proportional to ${\bf w}^2 $ dominates the potential, and the solution can be found
by the WKB approximation following \cite{Son:2002sd}. 
We  concentrate on the case $M_I =0$ in this section.

The Schr\"odinger equation (\ref{DiffLowT}) has a potential, which is positive 
when $0 < z < z_0$ and negative when $z_0  < z <1$, where 
\begin{align}
z_0 = \sqrt{\frac{{\bf s}^2}{1+ {\bf s}^2}} = 
\sqrt{\frac{ 2 M_R \omega + \vec k^2}{(M_R/(2b)+b \omega)^2+ \vec k^2 }}  \;. 
\end{align}
 Thus the solution of the equation (\ref{DiffLowT})
decays exponentially in the interval $0< z< z_0$ and oscillates in the interval $z_0< z< 1$.
Physically the particle has to tunnel from $z=0$ to $z=z_0$ before it can reach the horizon $z=1$.
The imaginary part of $G^R$ is proportional to the tunneling probability
\begin{align}
\text{Im}~ G^R \sim   \exp \left( -2 |{\bf w}| \int_0^{z_0} dz \sqrt{F(z)} \right) \;, 
\end{align}
and this integral is easy to evaluate for the following two limits :
\begin{itemize}
\item For small ${\bf s}$,%
\footnote{One might wonder that this condition invalidates the assumption $ {\bf q}^2, {\bf w}^2 \gg 1$. 
For ${\bf w} \geq {\bf q}$, one can evaluate the following differential equation 
\begin{align}
&g'' = \left({\bf q}^2 \tilde F  + \tilde H  \right) g  \;,  \qquad 
\tilde F =  \frac{(1-z^2) - z^2 \tilde {\bf s}^2 }{z \left(1-z^2\right)^2}\;, \qquad \tilde H = H\;.
\end{align}
where $\tilde {\bf s}^2 = 1/{\bf s} ^2 \geq 1$. 
Then the term proportional to ${\bf q}^2 $ dominates the potential, and the solution can be found
in a similar manner. 
}
we get $\int_0^{z_0} dz \sqrt{F(z)} =-\frac{\sqrt{2} \pi^{3/2} }{\Gamma \left(-\frac{1}{4} \right) 
\Gamma \left(\frac{7}{4} \right)}~({\bf {\bf s}^2})^{3/4} \;, $ 
and 
\begin{align}
\text{Im}~ G^R 
= \exp\left(-\frac{ 0.556}{b T} \cdot  \frac{ ( 2 M_R \omega + \vec k^2 )^{3/4}  }{(M_R/(2b)-b \omega)^{1/2}}  \right) \;. 
\label{LoglargeW}
\end{align}
\item For large ${\bf s}$, we get $\int_0^{z_0} dz \sqrt{F(z)} =\frac{2 \sqrt{\pi }  \Gamma\left(\frac{5}{4}\right)}{\Gamma\left(\frac{3}{4}\right)}\sqrt{{\bf s}^2}$, and thus  
\begin{align}
\text{Im} ~G^R
= \exp \left(   - 0.835 ~ \frac{\sqrt{ 2 M_R \omega + \vec k^2}}{b T} \right) \;. 
\label{LoglargeQ}
\end{align}
\end{itemize}
Thus, at low temperature, we obtain the generic behavior of the imaginary part of the 
correlation functions as an exponentially decaying one, $e^{-A / T}$. 
Specifically, the result for large ${\bf q}$, where $\text{Im} ~G^R 
\sim \exp \left(   - {\bf q} \right) $, is further confirmed by our numerical study in below.

\subsection{Numerical Results} \label{sec:numerics}

\begin{figure}[!th]
\begin{center}
\begin{tabular}{cc}
	 \includegraphics[width=0.45\textwidth]{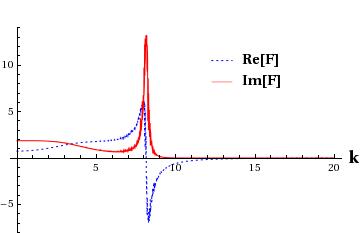} \qquad 
	 \includegraphics[width=0.45\textwidth]{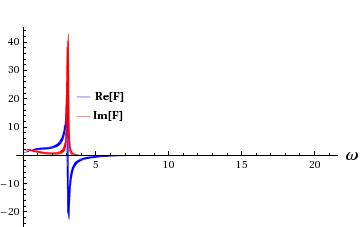}
 \end{tabular}
 \caption{Real and Imaginary parts of the scalar correlation function in terms of the 
 parameters $k$ with $\omega=0$ for left plot and $\omega$ with $k=0$ for right plot 
 for the fixed values $M=10, b=1$ and $T=1$. 
 There exist well defined peaks in imaginary part of the correlation function for both cases.}
\label{fig:NumBigMZeroMass2D}
\end{center}
\end{figure}

In this section we would like to present the numerical results of the correlation functions 
by solving the differential equation (\ref{MassiveEq}). 
For massless $m^2 =0$ case, the series solutions of the boundary and horizon are 
presented in appendix \ref{sec:SeriesSolm=0}. Massive case has also similar solutions 
with an additional parameter $m^2 L^2$, whose behaviors are briefly explained below.

\subsubsection{Description in terms of $\omega$ and $k$} \label{sec:omegakDesc}

For fixed $M=10, b=1$ and $T=1$, we present plots for the real and imaginary parts of the 
two-point correlation function in figure \ref{fig:NumBigMZeroMass2D}. The left plot is for 
the fixed $\omega = 0$. There exist a sharp peak at finite momentum and 
also a much broader peak at zero momentum $k=0$ for zero frequency.  
As we increase $\omega$ to the positive value, the peak moves toward to the $k=0$ and 
the peak becomes bigger and narrower, while it becomes smaller and broader as we decrease 
$\omega$ to the negative value. 
The right plot is for the fixed $k=0$. There also exists a sharp peak at finite 
value of $\omega$ for zero momentum. As $k$ increase, the sharp peak moves toward 
negative $\omega$ and becomes smaller and broader. Thus, overall, the sharp peak becomes broader and 
smaller as we increase $k$ and decrease $\omega$.    

\begin{figure}[!ht]
\begin{center}
\begin{tabular}{cc}
	 \includegraphics[width=0.7\textwidth]{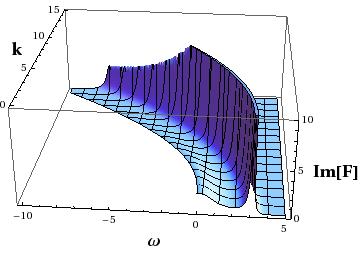}
 \end{tabular}
 \caption{Imaginary part of the scalar correlation function in terms of the 
 parameters $\omega, k$  for the fixed values $M=10, b=1$ and $T=1$. 
 They are even functions of $k$ and we take the parameter range as $-10 \leq \omega \leq 5, ~0 \leq k \leq 15$. 
 The plots are only drawn for the parameter range $2 M\omega + k^2 \geq 0$.}
\label{fig:NumBigMZeroMass3D}
\end{center}
\end{figure}

These behaviors can be checked in the figure \ref{fig:NumBigMZeroMass3D}. 
Note that we exclude the region which gives imaginary ${\bf q}$. 
There is another much broader peak at $(\omega, k)=(0,0)$, which can be identified as 
the point ${\bf q}=0$. This peak becomes smaller and broader 
along the line identified by ${\bf q} \sim 2 M \omega + k^2 =0$. 
This peak can be clearly seen in an alternative description below, section \ref{sec:QWDesc}. 
The above mentioned sharper peak seems to follow the broader peak line with some fixed distance, 
which can be described as $2M \omega + k^2 = \frac{M^2}{b^2}$. This peak is identified as 
the case where ${\bf w}=0$. Thus we confirm that these two peaks are related to the case when 
each of the last two terms in (\ref{MassiveEq}) has vanishing contributions. 
 
Let us comment about novel features of the scalar correlation functions for the
Schr\"odinger geometry.
First, we clearly have two peaks, of which the higher peak is
distinctively clear and
directly related to the unique property of the Schr\"odinger geometry,
the isometry $M$ and
associated chemical potential $\frac{1}{2b^2}$ of the spectator direction.
This, apparently, is analogous to the quasi-particle peak at the Fermi
surface in the case of spinor.
Second, the shape of our peaks in the $(\omega, k)$ space follows a curve
$2 M \omega + k^2 = const.$ due to the typical energy dispersion relation in
non-relativistic Schr\"odinger type, rather than relativistic ones
which would be
linear in the same parameters.

\subsubsection{Description in terms of ${\bf q}$ and ${\bf w}$} \label{sec:QWDesc}

While it is more physical to parametrize the physical properties of the correlation function 
in terms of $\omega, k$, some of its properties are more transparent in terms of the 
parameters, ${\bf w}$ and ${\bf q}$. Here we add some comments on them. 

\begin{figure}[!ht]
\begin{center}
\begin{tabular}{cc}
	 \includegraphics[width=0.65\textwidth]{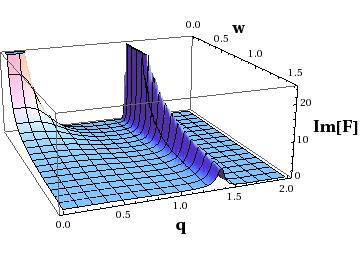} 
 \end{tabular}
 \caption{Imaginary part of the scalar correlation function in terms of the parameters ${\bf w}$ and ${\bf q}$.
 There are two peaks we observed in the previous plot, figure \ref{fig:NumBigMZeroMass3D}. }
 \label{fig:3DPlotQW}
\end{center}
\end{figure}

The typical shape of the correlator is depicted in figure \ref{fig:3DPlotQW}. 
We would like to identify the corresponding peaks we observed in section \ref{sec:omegakDesc}.
One of the peaks is located in ${\bf q} =0$, which is matched to the broad peak located in $\omega=k=0$. 
This is not exactly located at the origin but located in slightly non-zero value of the parameter ${\bf w}=\frac{M}{4\pi b^2 T}$, 
 due to the non-trivial parameter $M$ for the Schr\"odinger holography. 
The other peak is very sharp in terms of the parameter ${\bf q}$ and spread out in ${\bf w}$.
It is located at finite ${\bf q}^2 = \frac{M^2/b^2 + \vec k^2}{(2\pi b T)^2}$, 
we can identify this point near ${\bf w}=0$. 
As we increase the value of ${\bf w}$, the peak gets smaller and broader, but the 
peak location does not change much in ${\bf q}$.

Let us comment about the behavior of correlation function as we change the bare mass of the 
scalar $m^2 L^2$, which correspond to the change of the conformal dimension of the 
corresponding Field theory operator. 
For the range $-4 < m^2 L^2 < -3$, there is a single peak at ${\bf q}=0$ and 
at small ${\bf w}$, which spread out as ${\bf w}$ increases. As we increase $m^2 L^2$, the single 
peak moves to some positive value of ${\bf q}$ without being separated from the peak at 
$-{\bf q}$. Further increase of the bare mass brings the peak to the origin around at $m^2 L^2=0$. 
For positive mass range, $m^2 L^2 > 1$, there exist two separate peaks which are similar to 
the zero mass case in figure \ref{fig:3DPlotQW}, while the peaks are small compared to the 
zero mass case. 
With the numerical analysis, we find that massless $m^2 =0$ case is rather special and 
has a well defined isolated peaks at ${\bf q}=0$ and ${\bf w}=0$ and spreading as we change 
the parameters.

\begin{figure}[!ht]
\begin{center}
\begin{tabular}{cc}
	 \includegraphics[width=0.5\textwidth]{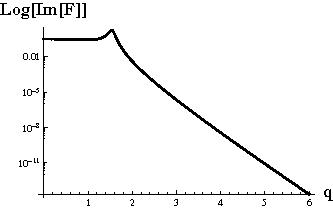}
 \end{tabular}
 \caption{A typical log plot of the imaginary part for ${\bf w} = 3$.}
 \label{fig:LogPlotLargeQ}
\end{center}
\end{figure}

Finally we would like to comment on the correlation function at large ${\bf q}$ and large ${\bf w}$ with 
${\bf q} \geq {\bf w}$. The figure \ref{fig:LogPlotLargeQ} represents a typical behavior in the region. 
We can check the exponentially decaying behavior we obtained with analytic analysis in equation (\ref{LoglargeQ}).

\subsection{Comments on Schr\"odinger Background}

In this section we consider the aging holography from the generalized Schr\"odinger background
\cite{Herzog:2008wg}\cite{Maldacena:2008wh}\cite{Adams:2008wt} generated through the null Melvin twist.
The action is given by 
\begin{align}
&S= \frac{1}{16\pi G} \int d^{5} x  \sqrt{-g}  \left( R - \frac{4}{3}  \partial_{\mu} \phi  \partial^{\mu} \phi  -\frac{1}{4} e^{-8\phi/3} F^2 - 4 A^2  -4 e^{ 2\phi /3 } \left( e^{2\phi} -4 \right) \right) \;, 
\end{align}
where $A$ and $F$ are a massive vector field and the associated field strength, and $\phi$ is a dilaton. The 
 black hole geometry associated with the action  was found as 
\begin{align}
ds^{2} &= r^{2} k^{-2/3}\left[ \left( \frac{1-h}{4 b^{2}} -r^{2} h \right) (dx^{+})^{2} + \frac{ b^2 r_H^{4}}{r^{4}} (dx^{-})^{2} -(1+h)dx^{+}dx^{-} \right]  \nonumber \\
& + k^{1/3} \left( r^{2} d\vec{y}^{2} + \frac{dr^{2}}{r^{2} h} \right) \;, 
\end{align}
where
\begin{align}
& \qquad h = 1- \frac{r_H^{4}}{r^{4}} \;, \qquad
k = 1+ \frac{b^2 r_H^{4}}{r^{2}} \;.\nonumber
\end{align}
The massive vector field and dilaton also have the non-trivial configurations as 
\begin{align}
&A=\frac{r^{2}}{k}\left( \frac{1+h}{2} dx^{+} -\frac{b^2 r_H^{4}}{r^{4}} dx^{-} \right)\;, \qquad
e^{\phi} =\frac{1}{\sqrt{k}} \;.\nonumber
\end{align}

Following the previous section, we use the local coordinate transformation (\ref{SingularTR}) 
to generate Aging black hole inherited from the Schr\"odinger background as
\begin{align}
ds^{2} &= r^{2} k^{-2/3}\left[ \left( \frac{1-h}{4 b^{2}} -r^{2} h \right) (dx^{+})^{2} + \frac{ b^2 r_H^{4}}{r^{4}} \left( d x^{-} +\frac{\alpha}{r}d r +\frac{\alpha}{2x^{+}}d x^{+}\right)^{2} \right.\nonumber\\
&\quad \left. -(1+h)dx^{+}\left( d x^{-} +\frac{\alpha}{r}d r +\frac{\alpha}{2x^{+}}d x^{+}\right) \right]
+k^{1/3} \left( r^{2} d\vec{y}^{2} + \frac{dr^{2}}{r^{2} h} \right)\;,
\end{align}
with the massive vector field
\begin{eqnarray}
A=\frac{r^{2}}{k} \left[ \left( \frac{1+h}{2} -\frac{\alpha b^2 r_H^{4} }{2x^{+}r^{4}} \right) dx^{+} -\frac{b^2 r_H^{4}}{r^{4}} dx^{-} - \frac{\alpha b^2 r_H^{4}}{r^{5}} dr \right] \;.
\end{eqnarray}
We can check that the metric reduces to that of \cite{Jottar:2010vp} when $r_{h} \rightarrow 0 $ and also
to that of \cite{Herzog:2008wg}\cite{Maldacena:2008wh}\cite{Adams:2008wt} when $\alpha \rightarrow 0$
by construction. One can make sure  that these configurations satisfy the Einstein equation.

We consider a probe free scalar on this background following the section \ref{sec:finiteTeq} with 
 the same equation of motion (\ref{probeEq}). Using the mode expansion 
\begin{align}
 \phi(r,x^{+},\vec{y}) &= \int \frac{d\omega}{2\pi} \frac{d^{2} k}{(2\pi)^{2}} \exp^{-i \vec{k}\cdot \vec{y} }  T_{\omega}(x^{+}) f_{\omega, \vec{k}}(r) \phi_{0}(\omega,\vec{k}) \;,
\end{align}
we can separate the $x^{+}$ dependent part of the differential equation with the same operator 
$\hat{D}$ given in equation (\ref{TimeOp}). The solution of the time dependent part is 
exactly the same as before, $T_{\omega}(x^{+}) = c_{1} \exp^{-i \omega x^{+}}(x^{+})^{\frac{i \alpha M}{2}}$.
This is one of the main observation for the aging backgrounds compared to the Aging in light-cone. 

Upon changing $z=\frac{r_{h}^{2}}{r^{2}}$ and 
$f_{\omega,\vec{k}}(z)=z^{- \frac{i \alpha M}{2} }\tilde{g}_{\omega,\vec{k}} (z)$, 
the $z-$dependent equation becomes 
\begin{align}
&\tilde g ''- \frac{1+ z^2 }{z (1-z^2) } \tilde g ' - \frac{(  m^2 k^{1/3}+M^{2}) }{4 z^2 (1-z^2)} \tilde g
+   \frac{{\bf w}^2 z }{ (1-z^2)^2} \tilde g
-  \frac{{\bf q}^2 }{z (1-z^2)} \tilde g = 0 \;,  \nonumber \\
&\text{where} \qquad {\bf w} =  \frac{M/(2b^2)- \omega }{2 \pi  T} \;, \qquad {\bf q}^2 =  \frac{2 M \omega + \vec k^2}{(2 \pi b T)^2}\;,
\label{SchrMassiveEq}
\end{align}
with the same temperature relation $r_H = \pi bT$ and with $L=1$ for the notational convenience. 
Again, this differential equation is identical to that of the Schr\"odinger backgrounds, 
confirming the general relations (\ref{FinalWaveFunction}), (\ref{AgingSchrCorrReal}) and 
(\ref{AgingSchrCorrIm}) in the Schr\"odinger black hole backgrounds. From these observations, 
we argue that the local coordinate transformation 
(\ref{CoordinateChange}) brings the same time dependent factor 
$\left( \frac{x^{+}_{1}}{x^{+}_{2}} \right)^{-\frac{ i \alpha M_R}{2}}$ for $M=M_R$ given in equation (\ref{AgingSchrCorrReal})
and $ \left(\frac{ x_2^+}{ x_1^+}\right)^{-\frac{\alpha M_I}{2}}$ given in equation (\ref{AgingSchrCorrIm}) to the 
correlation functions for the holographic aging phenomena. 

This equation (\ref{SchrMassiveEq}) is the same as the corresponding equation (\ref{MassiveEq}) of the 
AdS in light-cone except the following change 
\begin{align}
m^2 \quad \longrightarrow  \quad m^{2} ~ k(z)^{1/3} + M^{2} \;,
\end{align}
which is another main observation of this section. We would like to comment about the 
consequences of this change without detailed investigation of this differential equation. 
\begin{itemize}
\item In general, $k$ depends on the radial coordinate, which is  crucially different from the case of  
AdS in light-cone. In there, the corresponding term in equation (\ref{MassiveEq}) is constant and thus it is 
possible to consider the case $m^2=0$ consistently. On the other hand, it is not possible to 
ignore this term for the aging Schr\"odinger case in general with non-zero $M$. 
This seems to be the first non-trivial evidence that there exist differences between the two different 
geometric realizations of Schr\"odinger holography.  

\item $k$ is non-singular and monotonically increasing function 
from $k=1$ at the boundary to $k= 1+\pi^2 b^4 T^2$ at the horizon.  

\item This is the finite temperature generalization of the difference between the AdS in light-cone 
and the Schr\"odinger backgrounds, $m^2  \rightarrow m^2 + M^2$, which remains to be true at the boundary 
even at finite temperature. Thus we conclude that the differential equation of the Aging background 
is the same as that of the AdS in light-cone at the asymptotic region. 
\end{itemize}
It seems that the differential equation (\ref{SchrMassiveEq}) is similar to 
the massive case of Aging in light-cone given in (\ref{MassiveEq}). 
It remains to be seen the differences in detail with further investigations, which 
we postpone for the future.

\section{Conclusion and Outlook} 

In this paper we studied the time dependent aging systems by considering their dual gravity solutions with relevant symmetry, namely aging symmetry. We mainly focused on AdS in light-cone for the field theory with Schr\"odinger symmetry and its aging generalizations to incorporate the aging phenomena. Using these geometries, we study the two-point correlation functions of the dual scalar fields in aging system at zero temperature as well as   those in Schr\"odinger system at finite temperature. At zero temperature, the two-time correlation function of the dual scalar fields exhibit all the relevant features of the typical aging system. Among others, it shows that the older is the system the slower it relaxes. At finite, but low,  temperature with Schr\"odinger isometry, the analysis of momentum space correlation function exhibits the exponentially decaying behavior as a function of the inverse temperature.

One of the important features in the Schr\"odinger holography and also in aging holography is the existence of the spectator direction in the bulk geometry. One usually chooses an eigenvalue $M$ in that $x^-$ direction, which appears as an extra free parameter in the boundary conformal field theory. In order to deal with the time dependent, open and  dissipative system, it is tempting to relax the nature of this spectator or, more appropriately, internal direction and allow $M$ to be a  complex eigenvalue. Indeed, we could see that all the relevant features of the two-time correlation function in boundary aging system are revealed from the imaginary part of $M$.   

Another important point in this paper is the clear differences we found between Aging in light-cone and Aging backgrounds 
at finite temperature. They are the conformal dimensions of the scalar operators and the mass dependence of the 
differential equations, which can be understood simply from the differences in the effective mass, $m^2$ in the former and 
$m'^2 =m^2 k(r)^{1/3} + M^2$ for the latter. 
For AdS in light-cone, we evaluate two-point correlation functions in momentum space for massless scalar case, $m=0$, with finite $M$ 
using numerical method and analytic approaches for two special cases, low and high temperature limits.  
It seems to be possible to connect this case to read off the viscosity in AdS in light-cone with finite $M$ for general setup, 
while that was not possible in 
Schr\"odinger background due to the presence of $m'$ in the fluctuation of the gravity modes. 
Further systematic investigations are required to see the details, while preliminary analysis is presented in \ref{sec:viscosity}. 
Thus AdS in light-cone and its generalization to aging seems to be more attractive in their physical applications 
at finite temperature on top of the simplicity and well-defined boundary structure. 

Surely, we still need to clarify many issues to use the AdS/CFT correspondence for the time dependent field theory. Further extensive study of the aging system in this paper may shed  some lights on the study in this direction. One particular example is the thermodynamic properties of time dependent backgrounds, which are readily calculated in this setup and appropriate applications can be clarified in a controllable setting due to the large isometry and simple setup \cite{HJK2}.

\section*{Acknowledgments}

We would like to thank to J. Hartong, C. Hoyos, M. J\"{a}rvinen, E. Kiritsis, M. Lippert, 
D. Martelli, D. Minic, Y. Oz, C. Panagopoulos, J. Sonnenschein, J. Troost and D. Yamada
for discussions, comments and correspondence,
especially to M. J\"{a}rvinen for his careful illustrations of numerical calculations. 
BSK is grateful to the members, especially J. Sonnenschein, of the Raymond and Beverly Sackler Faculty of 
Exact Sciences at Tel Aviv University for their warm hospitality during his visit. 
We also thank to the referee who provided several critical comments.  
SH is supported in part by the National Research Foundation of Korea(NRF)  grant funded by the Korea government(MEST) 
with  the grant number  2009-0074518 and the grant number 2009-0085995 and by the grant number 2005-0049409 
through the Center for Quantum Spacetime(CQUeST) of Sogang University. JJ is supported by the National Research 
Foundation of Korea(NRF)  grant funded by the Korea government(MEST) with  the grant number 2009-0072755.
BSK is supported through excellent grant MEXT-CT-2006-039047 and also partially supported by  
a European Union grant FP7-REGPOT-2008-1-CreteHEP Cosmo-228644 and by PERG07-GA-2010-268246.

\appendix

\section{High Temperature Correlator : Next Order}\label{sec:HighTNext}

In section \ref{sec:HighTCorr}, we evaluate the scalar two-point correlation functions for the 
first non-trivial order. We evaluate the next order with the same setup. Here we summarize only the results.  
\begin{align} 
\tilde g_{\omega,\vec{k}} (z) &= c (1-z)^{i {\bf w}/2} \nonumber \\ 
&\times \left\{1+ \left(i\frac{ {\bf w}}{2} - {\bf q}^2\right) \ln \frac{1+z}{2} {\bf q}^4 \left(\frac{5\pi ^2}{24}- \ln\left[\frac{1-z}{2}\right] \ln\left[\frac{1+z}{2}\right]+\frac{1}{2} \ln[z] \ln[1+z] 
\right. \right. \nonumber \\
	& \left. \left.   \qquad \qquad \qquad \qquad  -\frac{1}{2} \text{PL}[2,1-z]+\frac{1}{2} \text{PL}[2,-z]-\text{PL}\left[2,\frac{1+z}{2}\right]\right)  \right. \nonumber \\
& \qquad \left. 
-\frac{1}{6} i{\bf q}^2 {\bf w} \left(\pi ^2-6 \ln[2]^2+\ln[64] \ln[1-z]
\right. \right. \nonumber \\
	& \left. \left.   \qquad \qquad \qquad \qquad 
-6 \ln\left[\frac{1-z}{2}\right] \ln[1+z]-6\text{PL}\left[2,\frac{1+z}{2}\right]\right) \right. \nonumber \\
& \qquad \left. +\frac{1}{24}{\bf w}^2 \left(-2 \pi ^2+15 \ln[2]^2+12 \ln[1-z] \ln\left[\frac{1+z}{2}\right] \right. \right.   \nonumber \\
	& \left. \left.   \qquad \qquad \qquad \qquad +3 \ln[1+z] (-2 \ln[8]+\ln[1+z])
+12\text{PL}\left[2,\frac{1+z}{2}\right]\right) \right\} \;, 
\end{align}
where 
\begin{align}
c= \frac{24}{\pi ^2 \left({\bf q}^2-i {\bf w}\right)^2+3 \left(8+\ln [2] \left(-4 {\bf q}^4 \ln [2]+{\bf q}^2 (8+4 i {\bf w} \ln [2])+{\bf w} (-4 i+{\bf w} \ln [8])\right)\right)} \;.
\end{align}
Where $\text{PL}$ is $\text{PolyLog}$. 
Following the standard prescription, we normalize the wave solution to be unity at $z=0$ 
and use the incoming boundary condition at $z=1$.  

From the first two non-trivial orders of the expansion, we can calculate the 
momentum correlator as 
 \begin{align}
	\frac{{\cal F}}{ 2 b^4 \pi ^4 L^{3} T^4} &=   - \frac{{\bf  q}^2}{z_B}  -i {\bf w} + {\bf  q}^2  + {\bf  q}^4+ \left(2 i{\bf w} {\bf  q}^2+{\bf w}^2 \right) \ln [2]-{\bf  q}^4 \ln [4 z_B]   + \cdots  \;.
	\label{MomentumCorrelatorForSmallWQ}
\end{align}
Note that this is similar to the relativistic results \cite{Son:2002sd} if we identify the parameters 
our ${\bf w}$ and ${\bf q}$ as relativistic frequency and momentum, respectively. 
But the physical meaning of these two parameters are very different and 
we develop several imaginary contributions by introducing complex $M$ as we explained 
in section \ref{sec:HighTCorr}. In particular, there is another non-trivial contribution 
$ {\bf w}^2 \ln [2]$ in the next order in addition to $ -i {\bf w}$ even in the case ${\bf q}=0$.

\section{Series Solutions with $m=0$} \label{sec:SeriesSolm=0}

Here we solve the differential equation (\ref{MassiveEq}) at the boundary and at the horizon
with appropriate expansions in each regions. These series solutions are used in the section \ref{sec:numerics}. 
It is also interesting to consider general cases with non-zero mass $m^2$, we would like to 
concentrate on $m=0$ in the differential equation (\ref{MassiveEq}).  
These series solutions are used to evaluate the radial part ${\cal F}$ of the correlator as
\begin{align}
{\cal F} = &\frac{b^4 \pi ^4 L^{5} T^4}{2 z^3} \left(  \frac{\tilde g_{\omega,\vec{k}} (z)}{\tilde g_{\omega,\vec{k}} (z_B)}  \right)^*  \frac{4 z^2 h_z }{L^2} \partial_z \left(  \frac{\tilde g_{\omega,\vec{k}} (z)}{\tilde g_{\omega,\vec{k}} (z_B)}  \right) \bigg |_{z \rightarrow z_B}  \;. 
\end{align}

At the boundary, we found a series solution upto 6th order of the expansion as 
\begin{align}
\tilde g_{B} = A \left(\sum_{n=0}^{5} a_n z^n + \sum_{n= 2}^{5} b_n z^n \ln z \right)+ B \left( \sum_{n=0}^{5} c_n z^{n+2} \right) \;.
\end{align}
We explicitly list here only the first four terms as
\begin{align}
& a_0 = 1\;, \quad a_1 =- {\bf q}^2 \;, \quad  a_2 =0 \;,\quad a_3 =\frac{1}{9} \left(-3 ~ {\bf q}^2+2 ~ {\bf q}^6-3 ~ {\bf w}^2\right)\;, \nonumber \\
& b_2 = -\frac{1}{2}  ~{\bf q}^4 \;, \quad  b_3 =-\frac{1}{6}  ~{\bf q}^6 \;, 
\nonumber \\
&c_0 = 1\;, \quad c_1 =\frac{ {\bf q}^2}{3}\;, \quad c_2 =\frac{1}{24} \left(12+ ~{\bf q}^4\right) \;,  \quad 
 c_3 = \frac{1}{360} \left(84  ~{\bf q}^2+ ~{\bf q}^6-24  ~{\bf w}^2\right) 
 \;. 
\end{align} 
This solution has two independent parameters, $A$ and $B$, 
which can be fixed by the solving the differential equation at the horizon using incoming boundary condition
and by interpolating the solution to the boundary.

At the horizon we have a similar series solution upto 6th order of the expansion
\begin{align}
\tilde g_{H} =   (1-z)^{\frac{i {\bf w}}{2}} \left( \sum_{n=0}^{5} d_n (1-z)^n \right)  \;,
\end{align}
with incoming boundary condition at the horizon as explained in the main body. The first four coefficients are 
\begin{align}
& d_0 = 1\;, \quad d_1 =\frac{2 i  {\bf q}^2+ {\bf w}}{4 i-4  {\bf w}}  \;, 
\nonumber \\ 
& d_2 =-\frac{4  {\bf q}^4+ {\bf q}^2 (8+8 i  {\bf w})+ {\bf w} \left(-4 i+6  {\bf w}+i  {\bf w}^2\right)}{32 \left(-2-3 i  {\bf w}+ {\bf w}^2\right)} \;,\nonumber \\
&d_3 = -\frac{8  {\bf q}^6+12  {\bf q}^4 (8+5 i  {\bf w})+6  {\bf q}^2 (2+i  {\bf w}) (8+ {\bf w} (8 i+ {\bf w})) }{384 (6-i  {\bf w} (-11+ {\bf w} (-6 i+ {\bf w})))}  \nonumber \\ &\qquad 
 -\frac{  {\bf w} (-48 i+ {\bf w} (108+(54 i-5  {\bf w})  {\bf w}))}{384 (6-i  {\bf w} (-11+ {\bf w} (-6 i+ {\bf w})))} 
\;.
\end{align}
These two solutions, near boundary and near horizon solutions, can be connected numerically. 
And the two-point correlation function is given by   
\begin{align}
\frac{{\cal F} }{4 b^4 \pi ^4 L^{3} T^4 } = \frac{B }{A}
 \end{align}
 in terms of various parameters we are interested in.

\section{Shear Viscosity}\label{sec:viscosity}

The finite temperature hydrodynamic correlation functions for the off-diagonal component of
the energy-momentum tensor $\langle T_{y_1}^{~y_2}~ T_{y_1}^{~y_2} \rangle$ were evaluated
for the Schr\"odinger background in \cite{Herzog:2008wg}\cite{Adams:2008wt}, while that of the
AdS in light-cone was done in \cite{Kim:2010tf}. It is interesting to recall that
the viscosity-entropy ratio for the strongly coupled non-relativistic plasma also satisfies the
universal bound, $\frac{\eta}{s} = \frac{1}{4\pi}$ 
\cite{Herzog:2008wg}\cite{Adams:2008wt}\cite{Kim:2010tf} for $\omega /T\ll 1$ with
vanishing parameters including $M$ as confirmed in section \ref{sec:HighTCorr}.
If we demand $M=\vec k=0$, the calculations are simple and effectively identical to those with the
wave equation for a massless scalar field of AdS$_5$ black hole.

Let us revisit the earlier treatments to evaluate the shear viscosity 
concentrating on Schr\"odinger geometry, which essentially used the scalar correlation functions,
as advertised in \cite{Kovtun:2003wp}. While it is straightforward to identify the correlation functions
of the energy-momentum tensor to those of scalar operators for the relativistic case with the vanishing
spatial momenta, there exist subtleties due to the internal direction in the non-relativistic
Schr\"odinger case. It is argued in \cite{Herzog:2008wg} that $M=0$ is required 
to evaluate the shear viscosity in the Schr\"odinger background.  The scalar operator
has the conformal dimension $\Delta = 2+ \sqrt{4+L^2 m^2 +M^2 }$ in the Schr\"odinger
backgrounds, while  the gravity modes have the conformal dimension $\Delta =4$. Thus it is required to
put $M=m=0$ to calculate the shear viscosity using the scalar operator, and thus restricted to the sector, $M=0$.

The story is different for the AdS in light-cone. As we mentioned in several places, we concentrate on the fixed sector
of $M$ and we would like to consider the fluctuation equation of the off-diagonal component as in \cite{Herzog:2008wg}
\begin{equation}
\delta g_{y_1}^{y_2} = e^{-i \omega x^+ + i M x^-} \Phi (r) \;,
\end{equation}
which satisfy the following differential equation
\begin{align}
\Phi'' + \frac{4+h}{r h} \Phi'
+ \frac{\left(M^2 + 4 b^4 \omega^2  \right) (1-h) -4 b^2  M\omega
(1+h)}{4 b^2 r^4 h^2 }  \Phi = 0 \;.
\label{MetricFluc}
\end{align}
Thus we are supposed to solve this differential equation directly to get the correlation functions
related to the metric fluctuations. This equation is nothing but the scalar differential equation
given in (\ref{MassiveEq}) with $m=0$, expressed in the $z$-coordinate. Now a crucial difference comes into play. Note that the
conformal dimension of a scalar operator is given by $\Delta = 2+ \sqrt{4+L^2 m^2}$. Thus we
can just restrict ourselves to the case $m=0$, which is already taken care of in (\ref{MassiveEq}). This made it possible to 
read off the viscosity, for general sector $M$, readily from the boundary two-point correlation of the scalar fields using 
the equation (\ref{MetricFluc}). 
This would be regarded as  a crucial difference between the AdS in light-cone and Schr\"odinger backgrounds
in the finite temperature case. We defer the systematic analysis for finite $M$ with full generality to the future. 

For $m=0$, being required by the conformal dimension of the energy momentum tensor, and $M=0$, 
being required by the condition ${\bf q}=0$ (this restriction would be lifted for general setup), 
we can read off the shear viscosity from the equation (\ref{ImFFfinal})
\begin{align}
   \eta = \frac{ b^4 \pi ^3 L^{3} T^3}{16 G}  \qquad \rightarrow \qquad \frac{\eta}{s} = \frac{1}{4\pi} \;,
\end{align} 
where we use the entropy formula from \cite{Kim:2010tf} and take into account the factor $K$ in front of 
the action (\ref{ScalarAction}). This is the advertised entropy-viscosity ratio of 
the strongly-coupled non-relativistic fluid advertised in \cite{Herzog:2008wg}\cite{Adams:2008wt}\cite{Kim:2010tf}. 
The result is the same as the relativistic counter part for $M=0$.

\end{document}